\documentclass[superscriptaddress,amssymb,amsmath,nobibnotes,aps,prd,eqsecnum,showpacs,
nofootinbib]{revtex4}
\usepackage{titlesec}
\usepackage{mathtools}
\usepackage{graphicx}

\usepackage{multirow}
\usepackage[usenames,dvipsnames,svgnames]{xcolor}

\usepackage[english]{babel}

\usepackage{amsmath}

\usepackage{amssymb}

\usepackage{amsfonts}

\usepackage{subfig}

\usepackage{dcolumn}
\usepackage{epsf}
\usepackage{enumerate}
\usepackage{hhline}
\usepackage{array}
\usepackage{tabularx}
\usepackage{appendix}
\usepackage[colorlinks=true,
            linkcolor=blue,
          urlcolor=gray,
            citecolor=red]{hyperref}
\usepackage{epstopdf}
\newcommand{\be}{\begin{equation}}

\newcommand{\ee}{\end{equation}}

\newcommand{\bea}{\begin{eqnarray}}

\newcommand{\eea}{\end{eqnarray}}

\newcommand{\beaa}{\begin{eqnarray*}}

\newcommand{\eeaa}{\end{eqnarray*}}

\newcommand{\nn}{\nonumber \\}

\begin{document}


\title[Bouncing solutions in $f(T, T_G)$ gravity...]{
Cosmological bouncing solutions in extended teleparallel gravity theories}
\author{\'Alvaro de la Cruz-Dombriz}
\affiliation{Cosmology and Gravity Group, University of Cape Town, 7701 Rondebosch, Cape Town, South Africa} 
\affiliation{Department of Mathematics and Applied Mathematics, University of Cape Town, 7701 Rondebosch, Cape Town, South Africa}

\author{Gabriel Farrugia}
\affiliation{Department of Physics, University of Malta, Msida, MSD 2080, Malta}
\affiliation{Institute of Space Sciences and Astronomy, University of Malta, Msida, MSD 2080, Malta}

 \author{Jackson Levi Said} 
\affiliation{Department of Physics, University of Malta, Msida, MSD 2080, Malta}
\affiliation{Institute of Space Sciences and Astronomy, University of Malta, Msida, MSD 2080, Malta}
 
\author{Diego S\'aez-Chill\'on G\'omez}
\affiliation{Institut de Ci\`{e}ncies de l'Espai (ICE, CSIC), Carrer de Can Magrans s/n, Campus UAB, 08193 Bellaterra (Barcelona), Spain}
\affiliation{Institut d'Estudis Espacials de Catalunya (IEEC), 08034 Barcelona, Spain}


\begin{abstract}
In the context of extended Teleparallel gravity theories with a 3+1 dimensions Gauss-Bonnet analog term, we address the possibility of these theories reproducing several well-known cosmological bouncing scenarios in a four-dimensional Friedmann-Lema\^itre-Robertson-Walker geometry. We shall study which types of gravitational Lagrangians are capable of reconstructing bouncing solutions provided by analytical expressions for symmetric, oscillatory,  superbounce, the matter bounce and singular bounce. Some of the Lagrangians discovered are both analytical at the origin having Minkowski and Schwarzschild as vacuum solutions. All these results open the possibility up for such theories to be competitive candidates of extended theories of gravity in cosmological scales.

\end{abstract}

\maketitle

\section{Introduction}
\label{S1}
The appearance of cosmological bouncing scenarios has attracted much attention in latest years due to its power to avoid the unnaturalness of our Universe to be created from a big bang initial singularity. In such scenarios, the Universe contracts until reaching a minimal non-zero radius, bounces off and then expands ({\it c.f.} \cite{Brandenberger:2016vhg} and references therein for a recent thorough review on the subject), similarly to the so-called ekpyrotic scenario \cite{Khoury:2001wf}. Apart from the possibility of preventing the initial cosmological singularity, the so-called 
big bounce cosmologies have been shown to provide competitive scenarios to the standard inflationary paradigm~\cite{Mukhanov:2005sc}-\cite{Bamba:2015uma} 
and in some realisations, such as the so-called matter bounce scenario, 
to generate a nearly scale-invariant power spectrum as in usual inflationary models   \cite{Brandenberger:2012zb}-\cite{Barragan:2009sq}.
\\
~

As such, 
bouncing solutions in the context of gravitational theories beyond the Einsteinian paradigm have also drawn some attention in recent literature. 
Firstly, the idea of ekpyrotic/cyclic cosmologies were analysed in the framework of $f(R)$ gravities in Ref.~\cite{Nojiri:2011kd}. Related works on bounce cosmology reconstruction from scalar-tensor $f(R)$ theories can be found in 
\cite{Bamba:2013fha, Odintsov:2014gea}.
Other recent proposals such as the unimodular $f(R)$ gravity was studied in \cite{Nojiri:2016ygo} where the authors studied 
well-known cosmological bouncing models and investigated which era of the whole bouncing model is responsible for the cosmological perturbations.
Also, a seminal reference was \cite{Odintsov:2015uca} where the authors investigated the superbounce and the loop quantum cosmological ekpyrosis bounce for 
 $f(R)$, $f(G)$ and $f(T)$ gravity theories, showing the qualitative similarity of the different effective gravities realising the two bouncing cosmologies mentioned above. Moreover, by 
performing a linear perturbation analysis, it was shown that the obtained solutions are conditionally or fully stable.
Also in $f(T)$ extended teleparallel gravity authors in \cite{Cai:2011tc} focused on the simplest version of a matter bounce and studied  the scalar and tensor modes of subsequent cosmological perturbations. Results showed that scalar metric perturbations lead to a background-dependent sound speed, which might be distinguishable from the Einsteinian prediction, and a scale-invariant primordial power spectrum, which is consistent with cosmological observations. Indeed, one can infer that extensions of Teleparallel gravity reach a wide and rich family of solutions in the context of cosmology \cite{Linder:2010py}. In addition, some alternative formulations of Teleparallel gravity where the Palatini approach is applied, show some interesting properties when dealing with the boundary terms in the Euclidean action \cite{BeltranJimenez:2017tkd}.
\\

In the present work we shall investigate several well-established bouncing scenarios in the frame of extended teleparallel gravity theories with non-vanishing boundary terms, dubbed $f(T,T_G)$ theories, where an analog of the Gauss-Bonnet invariant is assumed in the framework of Teleparallel gravity \cite{Kofinas:2014owa}. The existence of cosmological solutions have already been studied in such theories, where some reconstruction methods were implemented (see Ref.~\cite{Kofinas:2014daa} and {\it c.f.} \cite{delaCruz-Dombriz:2017lvj} for a thorough review on the existence of cosmological solutions in such theories). 
Also static spherically symmetric solutions and its relation with other extensions of TE-GR have been analysed \cite{Bahamonde:2016kba}. Thus, we shall use the reconstruction method for $f(T,T_G)$ theories to realise such cosmological bouncing scenarios.
In particular, we shall apply this method to bouncing cosmologies in spatially flat four-dimensional Friedmann-Lema\^itre-Robertson-Walker geometries to paradigmatic bouncing solutions, 
such as the symmetric bounce \cite{Bamba:2013fha}; 
an oscillatory bouncing solution  where the universe oscillates through a series
of expansions and contractions \cite{Tolman, Steinhardt:2001st, Khoury:2003rt}; 
a generic power-law bounce 
which has been for instance studied in the context of  modified Gauss-Bonnet gravity \cite{Bamba:2014mya} and  loop quantum cosmology scenarios  \cite{Haro:2015oqa, Ranken:2012hp}; 
the superbounce  \cite{Koehn:2013upa, Odintsov:2015uca, Oikonomou:2014yua}; 
%
the matter bounce scenario \cite{Brandenberger:2012zb}-\cite{Barragan:2009sq}, also dubbed critical density bouncing,  
which naturally arises in loop quantum cosmology scenarios \cite{Ashtekar:2011ni}-\cite{Bojowald:2008ik}  
and provides a viable alternative scenario to inflation compatible with Planck data,
and finally the so-called singular bounce  \cite{Barragan:2009sq}, \cite{Odintsov:2015ynk}-\cite{Oikonomou:2015qha} in which the Hubble radius is infinite as $t\rightarrow-\infty$ and gradually decreases until a minimal size, but near the
bouncing point ($t=0$) it increases and blows up at exactly the
bouncing point. In this latter case, after the bouncing point the Hubble radius
eventually decreases gradually. This is different in comparison to other
bouncing cosmologies, and this can be seen by comparing
directly the behavior of the Hubble radius in Fig. \ref{Model1}.\\

For the sake of clarity, further technical details about each bouncing scenario shall be provided in upcoming sections.
Moreover,  in the bulk of the article we shall show that these bouncing solutions can be obtained in both 
universes filled with one standard fluid provided with a constant equation of state and, when possible, in vacuum configurations. Thus, our results show that within this class of theories bounce realisations do not rely on the existence of extra 
matter fields nor on the existence of fluids with an equation of state which violates the 
 null energy condition as it is the case in other bouncing scenarios \cite{Bouncings_others}. The types of gravitational actions analysed along the paper are based on the idea of extending Teleparallel gravity in such a way that the corresponding Lagrangians are constructed as separable (or multiplicative) additional terms, which perturbatively (depending on the extra parameters in the Lagrangians and the involved exponents) can be negligible in some scales but relevant in others  (cosmological). \\


The paper is organised as follows: in Section \ref{S2} we shall briefly remind the general features of the $f(T,T_G)$ gravity theories and the state of the art within this class of extended theories of gravity. There we shall provide the key equations to consider so the reconstruction mechanism can be performed. 
 In the following sections, we shall briefly discuss the main features of the
bouncing models to be studied and determine the 
$f(T,T_G)$ gravity theories capable of realising such cosmologies.
%
%
%
%
%
%
%
%
Thus, in Sec. \ref{model1} we shall discuss the reconstruction of the symmetric bounce.
 Then Section \ref{model2} addresses the same issue when the desired model to be reconstructed is a paradigmatic oscillatory bounce solution when paremeterised as a squared sine function.
 Finally, Sections \ref{model3}, \ref{model4} and \ref{model5} are devoted to studying the possibility of reconstruction of superbounce, matter and singular bounce solutions respectively.
We conclude the paper by giving our conclusions in Section \ref{Conclusions}. 
At the end of the paper, the scale factor, the Hubble
parameter and the torsion scalar are depicted in Fig. \ref{Model1} for a particular set of the free parameters for the five bouncing models under consideration. The
bouncing character of the solutions is clearly shown as well as the possible singularities that may occur.
\newline

Throughout the paper we shall follow the following conventions: the Weitzenb\"ock connection as defined in Sec. \ref{S2} will be denoted by $\tilde{\Gamma}^{\alpha}_{\mu\nu}$. $D_{\mu}$ shall represent  the covariant derivative with respect to the usual Levi-Civita connection $\Gamma^{\alpha}_{\mu\nu}$.
%
Greek indices such as $\mu, \nu...$ shall refer to spacetime indices whereas latin letters $a, b, c...$ refer to the tetrads indices associated to the tangent space. 

\section{$f(T,T_G)$ theories}
\label{S2}

Teleparallel gravities can be expressed  by defining the mathematical objects known as vierbeins  $e_{a}(x^{\mu})$, 
\be
{\rm d}x^{\mu}=e_{a}^{\;\;\mu}\omega^{a}\; , \quad \omega^{a}=e^{a}_{\;\;\mu}{\rm d}x^{\mu}\; ,
\label{1.1}
\ee
which relate the spacetime of a manifold with its the tangent space at every point $x^{\mu}$.
\begin{eqnarray}
{\rm d}s^{2} &=&g_{\mu\nu}{\rm d}x^{\mu}{\rm d}x^{\nu}=\eta_{ab}\omega^{a}\omega^{b}\label{1}\; ,
\label{1.1a}
\end{eqnarray} 
where $\eta_{ab}={\rm diag}(-1,1,1,1)$ holds for the Minkowskian metric. In addition, the tetrads accomplish the following properties:
\be
 e_{a}^{\;\;\mu}e^{a}_{\;\;\nu}=\delta^{\mu}_{\nu}\ , \quad e_{a}^{\;\;\mu}e^{b}_{\;\;\mu}=\delta^{b}_{a}\ .
 \ee
The theory is constructed as a gauge theory of the translation group, leading to the so-called  Weitzenb\"{o}ck connection, defined as:
\begin{eqnarray}
\tilde{\Gamma}^{\alpha}_{\mu\nu}=e_{a}^{\;\;\alpha}\partial_{\nu}e^{a}_{\;\;\mu}=-e^{a}_{\;\;\mu}\partial_{\nu}e_{a}^{\;\;\alpha}\label{co}\; ,
\label{WC}
\end{eqnarray}
Whereas the Riemann tensor becomes null under this connection, torsion does not vanish, such that the torsion scalar is defined as:
\begin{eqnarray}
T=T^{\alpha}_{\;\;\mu\nu}S^{\;\;\mu\nu}_{\alpha}=\frac{1}{4}T^{\lambda}_{\;\;\;\mu\nu}T_{\lambda}^{\;\;\;\mu\nu}+\frac{1}{2}T^{\lambda}_{\;\;\;\mu\nu}T_{\;\;\;\;\;\lambda}^{\nu\mu}-T^{\rho}_{\;\;\;\mu\rho}T_{\;\;\;\;\;\nu}^{\nu\mu}\, .
\label{scalar-torsion}
\end{eqnarray}
where the torsion tensor is given by:
\begin{eqnarray}
T^{\alpha}_{\;\;\mu\nu}&=&\tilde{\Gamma}^{\alpha}_{\mu\nu}-\tilde{\Gamma}^{\alpha}_{\nu\mu}=e_{a}^{\;\;\alpha}\left(\partial_{\nu} e^{a}_{\;\;\mu}-\partial_{\mu} e^{a}_{\;\;\nu}\right)\ . 
\label{tor}
\end{eqnarray}
and
\begin{eqnarray}
S_{\alpha}^{\;\;\mu\nu}&=&\frac{1}{2}\left( K_{\;\;\;\;\alpha}^{\mu\nu}+\delta^{\mu}_{\alpha}T^{\beta\nu}_{\;\;\;\;\beta}-\delta^{\nu}_{\alpha}T^{\beta\mu}_{\;\;\;\;\beta}\right)\label{s}\;,
\end{eqnarray}
Here the contorsion is given by the difference between the Weitzenb\"ock  and the Levi-Civita connection:
\be
K^{\alpha}_{\;\; \mu\nu}= \tilde{\Gamma}^{\alpha}_{\mu\nu}-\Gamma^{\alpha}_{\mu\nu}=\frac{1}{2}\left(T_{\mu\;\;\ \nu}^{\;\; \alpha}+T_{\nu\;\;\ \mu}^{\;\; \alpha}-T_{\;\; \mu \nu}^{\alpha}\right)\;,
\label{contor2}
\ee
Thus the gravitational action for TEGR is solely given by the torsion scalar (\ref{scalar-torsion}), 
\begin{eqnarray}
\label{action}
S_G=-\frac{1}{2\kappa^2}\int e\ T\ {\rm d}^4x\; ,
\label{TeleAction}
\end{eqnarray}
where $\kappa^2=8\pi G_{N}$, $G_N$ the usual gravitational constant, and $e=$ det $\left(e^{a}_{\;\;\mu}\right)$. This action is equivalent to the Einstein-Hilbert action, since the relation of the torsion scalar and the Ricci curvature is given by
\be
R\,=\,-T-2D_{\mu}T^{\nu\mu}_{\;\;\;\;\;\nu}\ .
\label{RS}
\ee
Here the last term is a total derivative and can be dropped out of the action. However, any non-linear function of the torsion scalar will not be equivalent to $f(R)$ gravity as shown in Eq. (\ref{RS}). \\

Recently, the analog to the Gauss-Bonnet term with the Weitzenb\"{o}ck connection was found by using the above expression:
\be
G=T_G+B_G\ ,
\label{TGGB}
\ee
where the Gauss-Bonnet invariant is defined as:
\be
G=R_{\mu\nu\lambda\sigma}R^{\mu\nu\lambda\sigma}-4R_{\mu\nu}R^{\mu\nu}+R^2\ .
\label{GBterm}
\ee
And the second term in (\ref{TGGB}) is a total derivative, such that $T_G$ can be expressed as follows \cite{Kofinas:2014owa}:
\begin{eqnarray}
T_G&=&\left(K^{\alpha}_{\;\;\gamma\beta}K^{\gamma\lambda}_{\;\;\;\;\rho}K^{\mu}_{\;\;\epsilon\sigma}K^{\epsilon\nu}_{\;\;\;\;\varphi} -2K^{\alpha\lambda}_{\;\;\;\;\beta}K^{\mu}_{\;\;\gamma\rho} K^{\gamma}_{\;\;\epsilon\sigma}K^{\epsilon\nu}_{\;\;\;\;\varphi}\right. \nn
&&+\left. 2K^{\alpha\lambda}_{\;\;\;\;\beta}K^{\mu}_{\;\;\gamma\rho}K^{\gamma\nu}_{\;\;\;\;\epsilon}K^{\epsilon}_{\;\;\sigma\varphi}+2K^{\alpha\lambda}_{\;\;\;\;\beta}K^{\mu}_{\;\;\gamma\rho}K^{\gamma\nu}_{\;\;\;\;\sigma,\varphi}\right)\delta^{\beta\rho\sigma\varphi}_{\alpha\lambda\mu\nu}.
\label{TG}
\end{eqnarray}
Hence, any linear action on $T_G$ leads to a total derivative, as in the metric case. Nevertheless, beyond the linear order the equivalence is broken. Here we are focusing on theories containing in the action such type of functions beyond the linear order on $T_G$,
\begin{eqnarray}
S=S_G+S_m=\int e\left(\ f(T,T_G)\ +2\kappa^2\mathcal{L}_m\right){\rm d}^4x\; .
\label{fttGaction}
\end{eqnarray}
By assuming a spatially flat FLRW metric, $T$ and $T_G$ can be expressed in terms of the Hubble parameter as follows
\begin{eqnarray}
T=6H^2\;\;\;;\;\;\; T_G=24H^2(\dot{H}+H^2)\,.
\label{T_TG_expressions}
\end{eqnarray}
Note that $T_G$ coincides with its GR counterpart, $G$, that is when assuming a spatial flatness. Then, the FLRW equations yield \cite{Kofinas:2014daa}
\begin{eqnarray}
& &f - 12H^2 f_T - T_G f_{T_G} +24H^3\dot{f}_{T_{G}}\,=\,2\kappa^2\rho_m\,, \label{Friedmann1} \\ 
& &f-4\left(3H^2+\dot{H}\right)f_T-4H\dot{f}_{T}-T_{G} f_{T_{G}}+\frac{2}{3H}T_{G} \dot{f}_{T_{G}}+8H^2\ddot{f}_{T_{G}}\,=\,-2\kappa^2p_m\, .\label{Friedmann2} \nonumber\\
&&
\end{eqnarray}
Here we have assumed the standard definition for the energy-momentum tensor $\mathcal{T}^{\;\nu}_{\mu}=\frac{e_{a}^{\;\;\nu}}{e}\frac{\delta\mathcal{L}_\mathrm{m}}{\delta e_{a}^{\;\;\mu}}$, together with the assumption of a perfect fluid. Combination of the previous equations leads to the usual conservation of the energy-momentum tensor. Thus, by using the above tools, we are considering several types of bouncing solutions in the next sections and some classes of Lagrangians are reconstructed.

\section{Bouncing cosmology I: Exponential evolution}
\label{model1}

\begin{figure}[h!]
\centering
\includegraphics[scale=0.5]{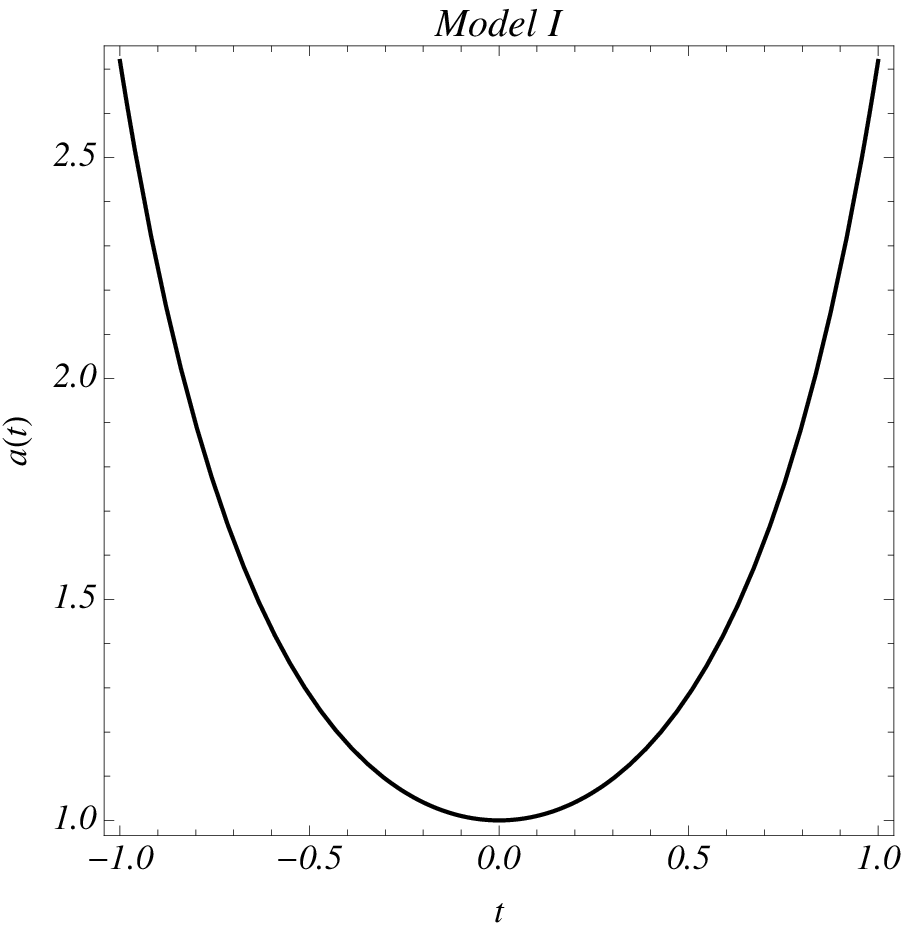}
\includegraphics[scale=0.5]{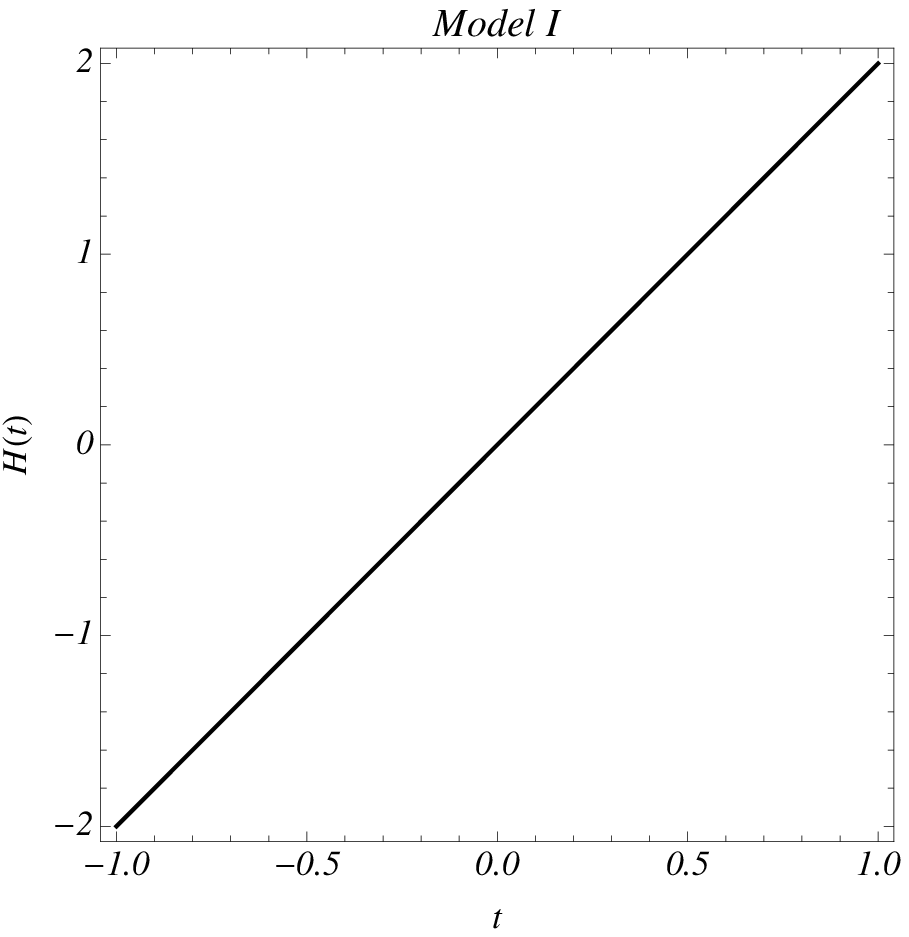}
\includegraphics[scale=0.5]{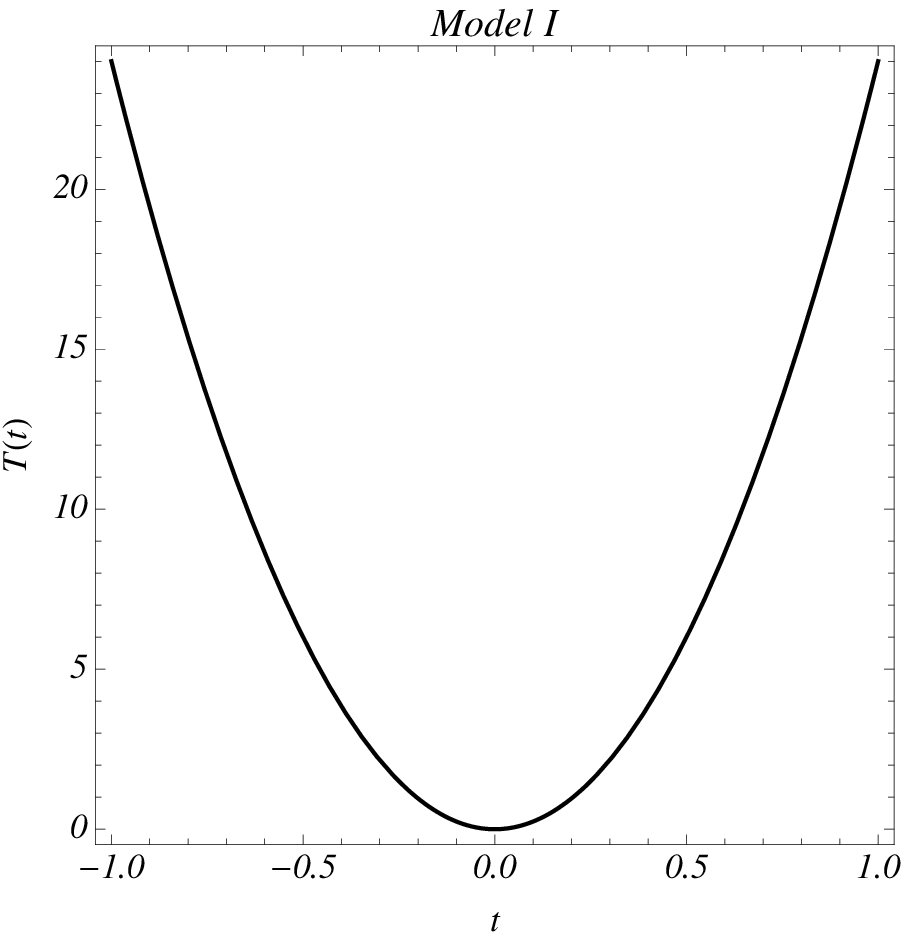}
\includegraphics[scale=0.5]{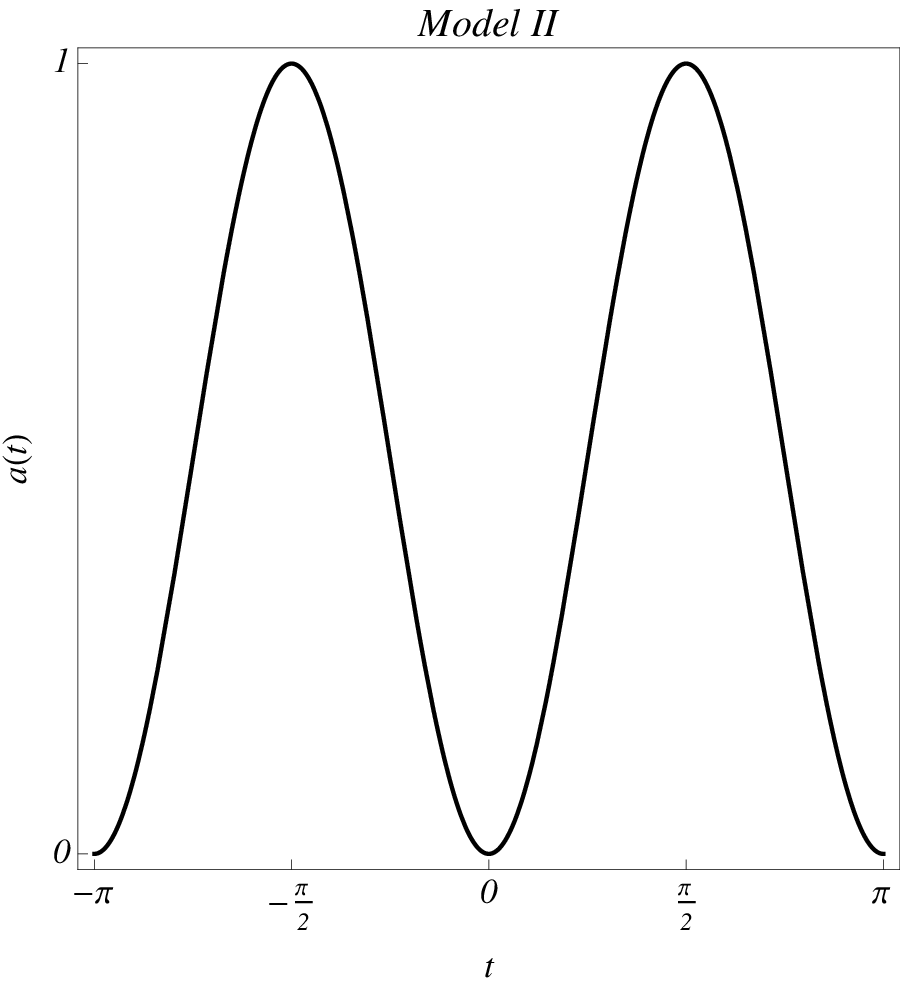}
\includegraphics[scale=0.5]{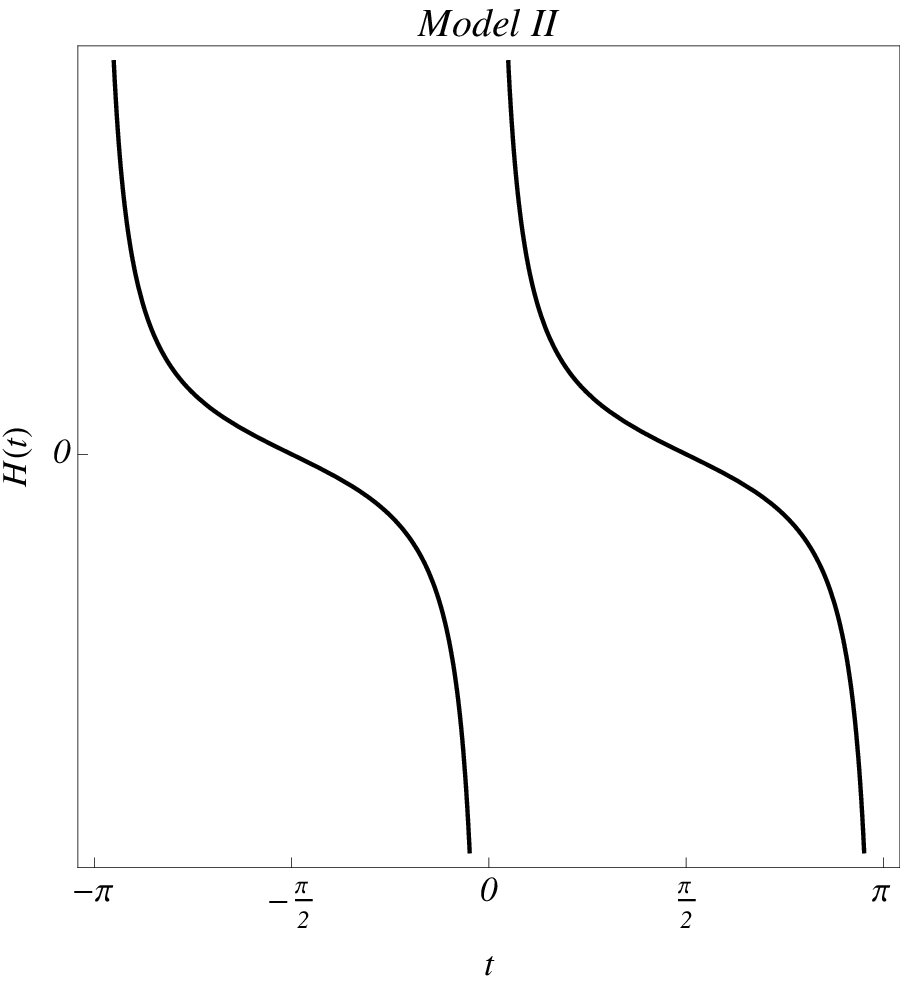}
\includegraphics[scale=0.5]{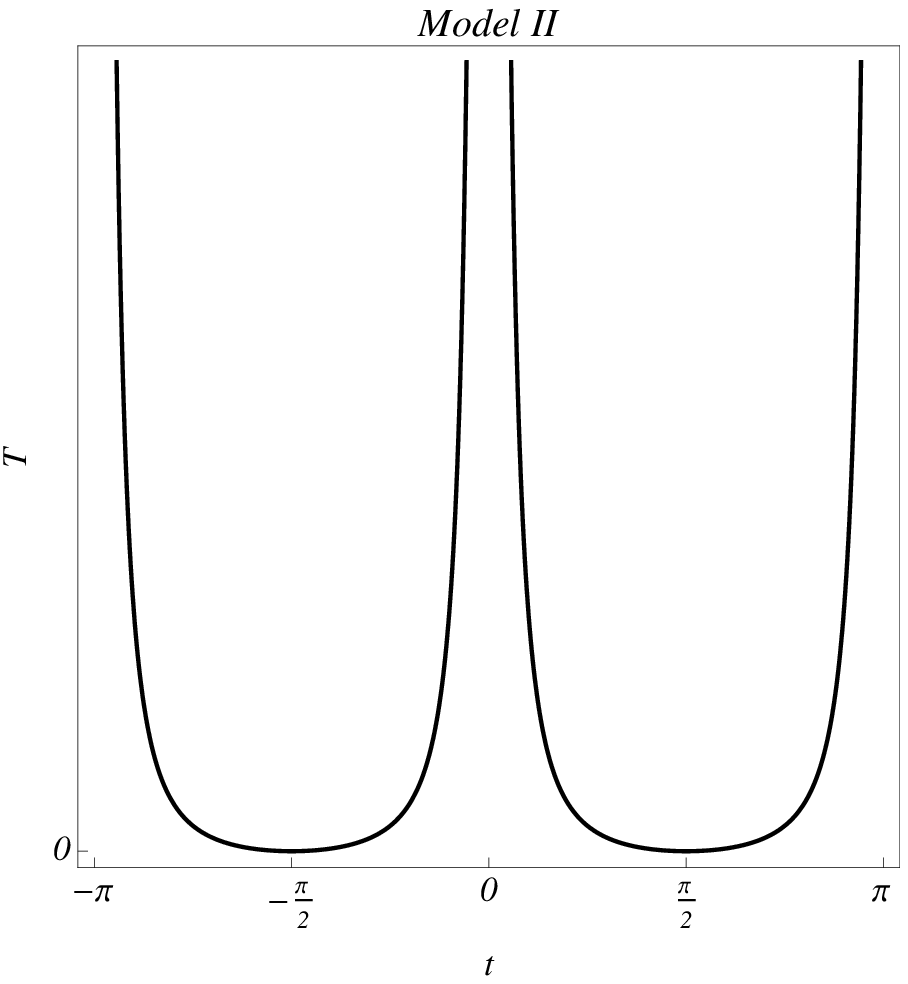}
\includegraphics[scale=0.5]{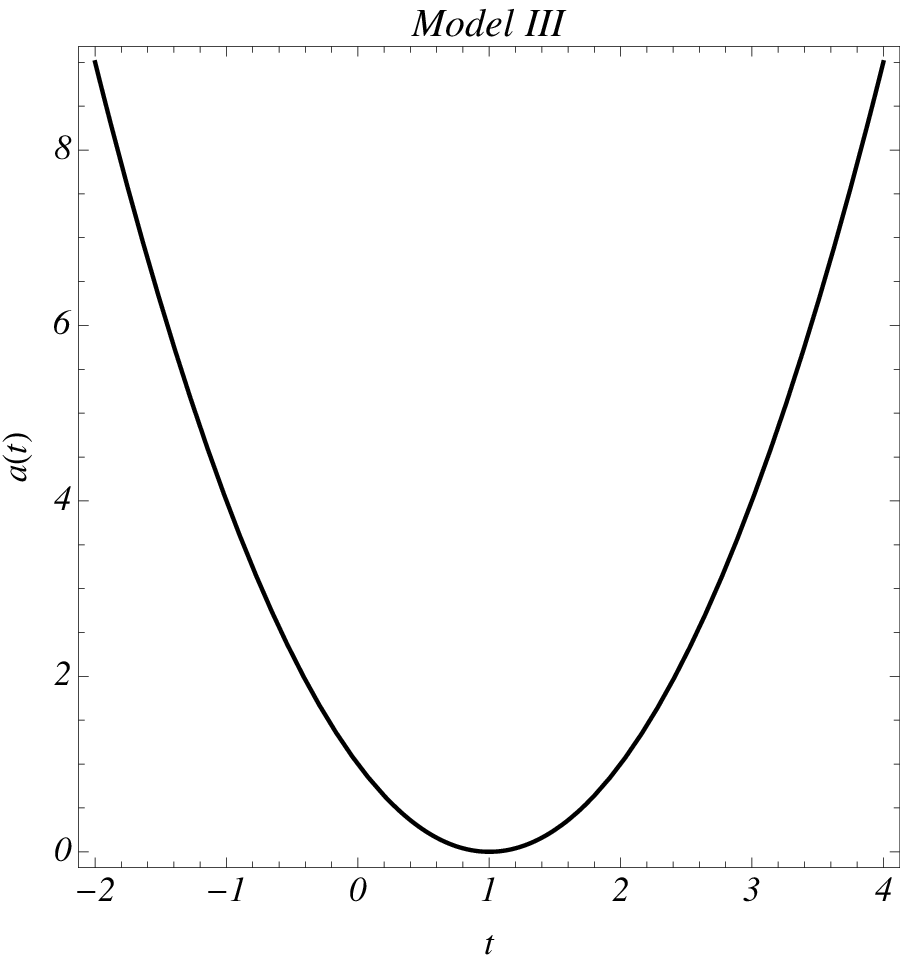}
\includegraphics[scale=0.5]{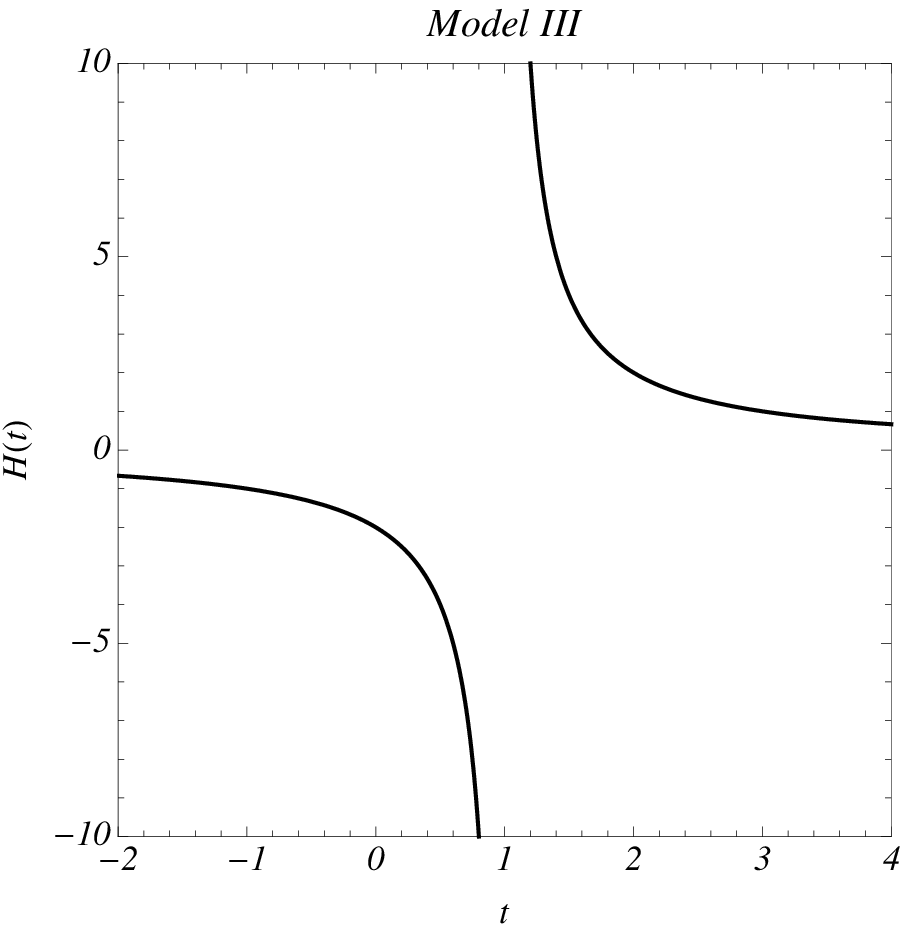}
\includegraphics[scale=0.5]{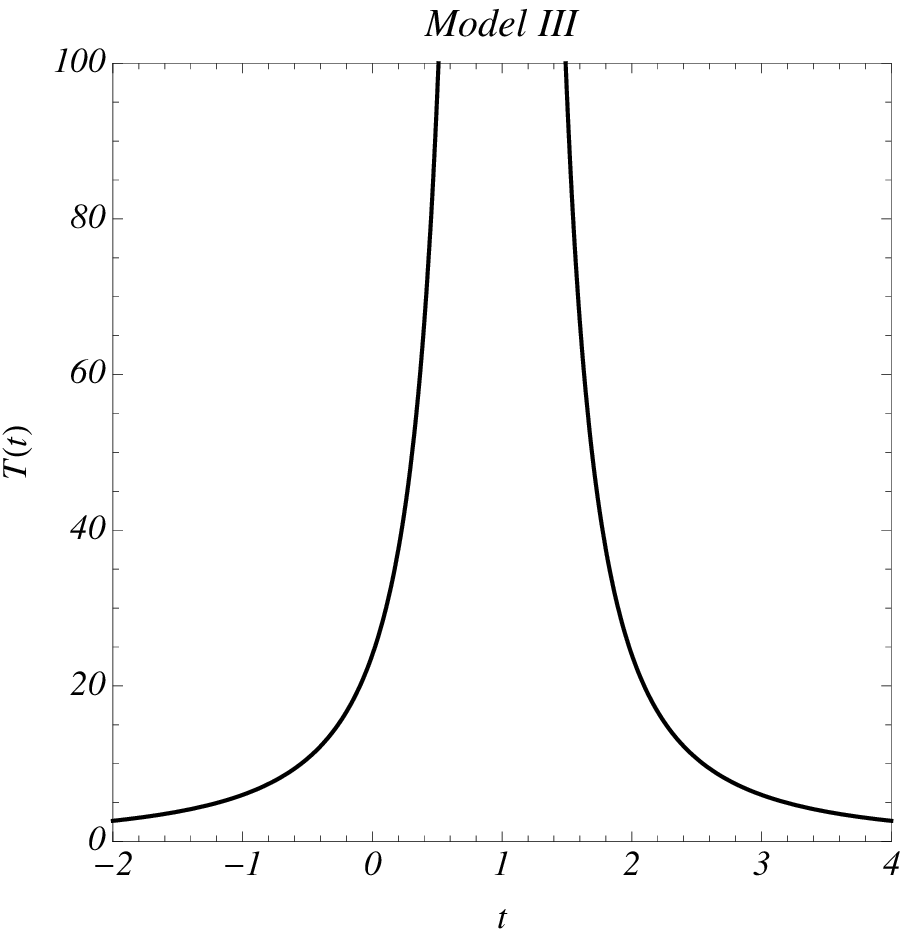}
\includegraphics[scale=0.5]{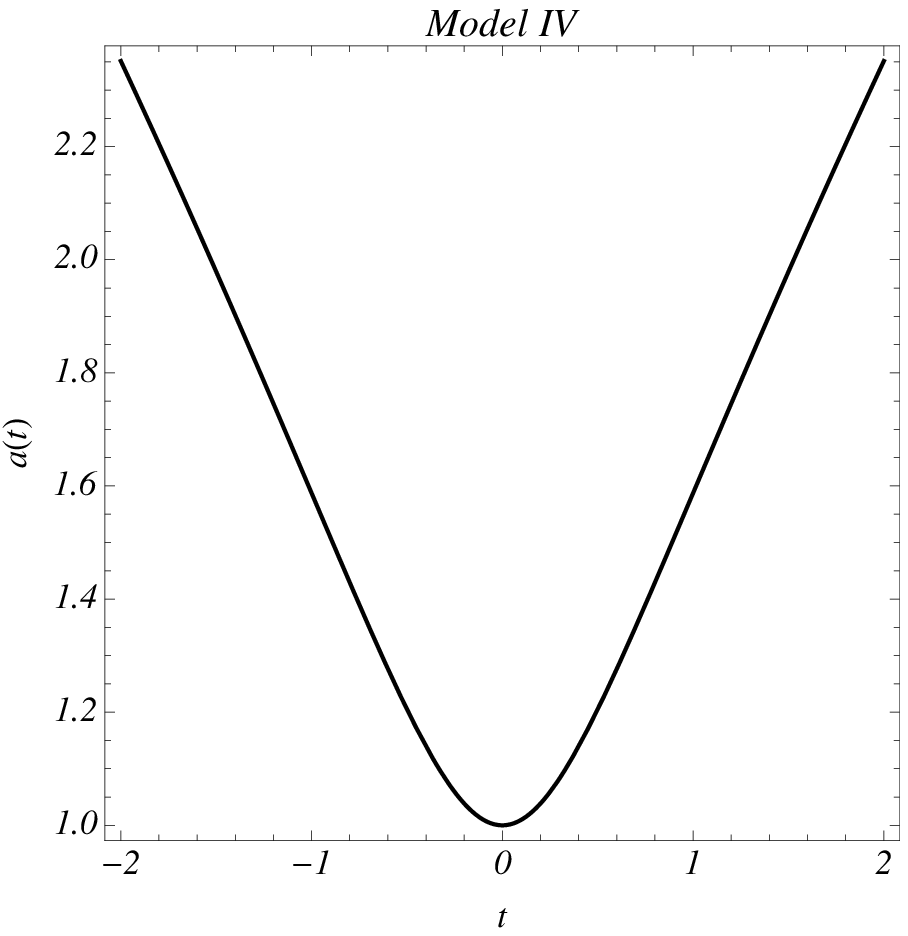}
\includegraphics[scale=0.5]{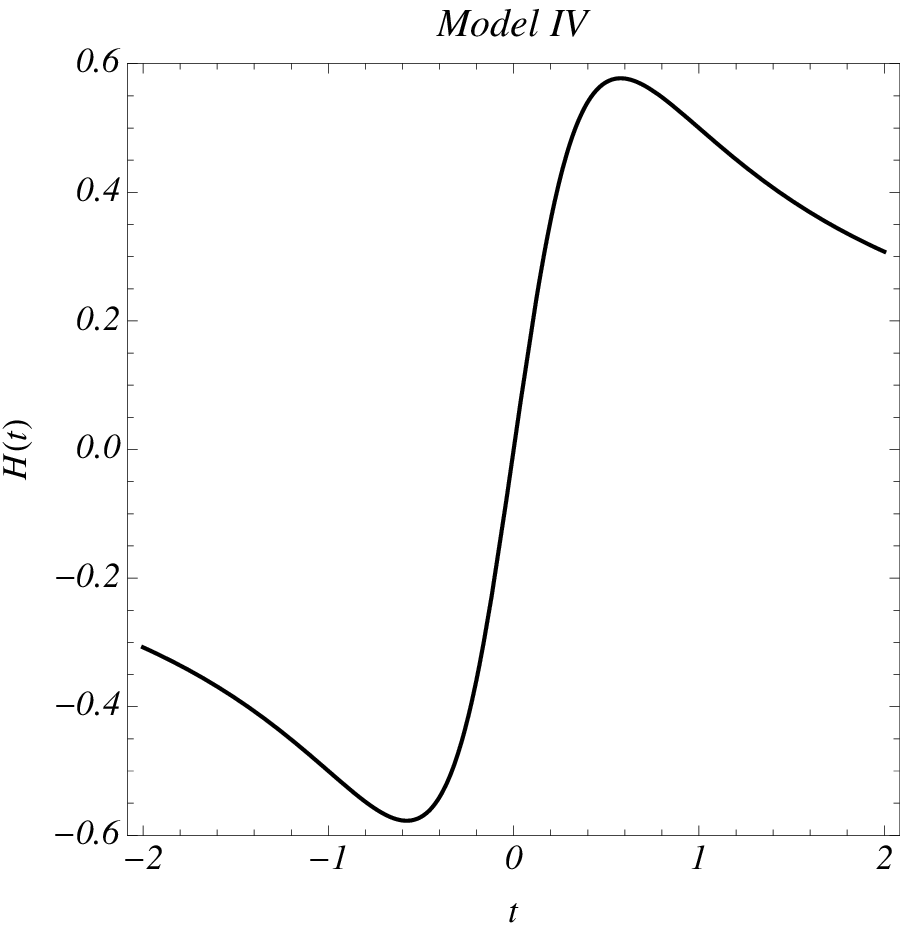}
\includegraphics[scale=0.5]{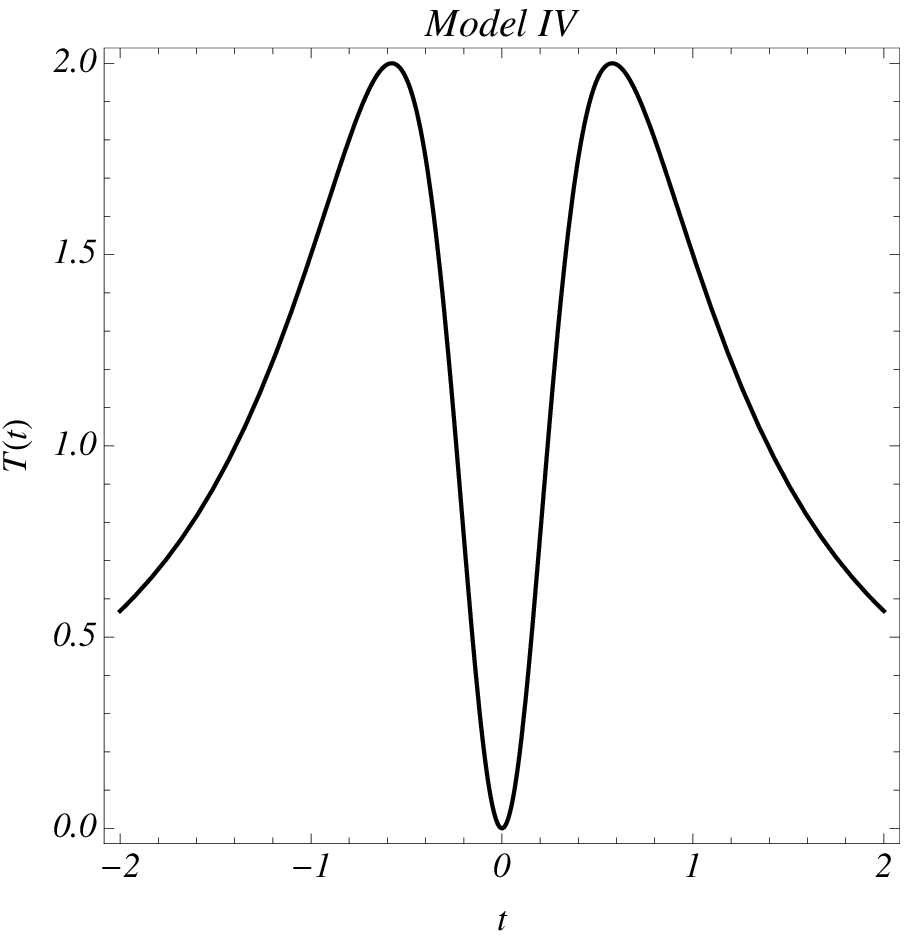}
\includegraphics[scale=0.5]{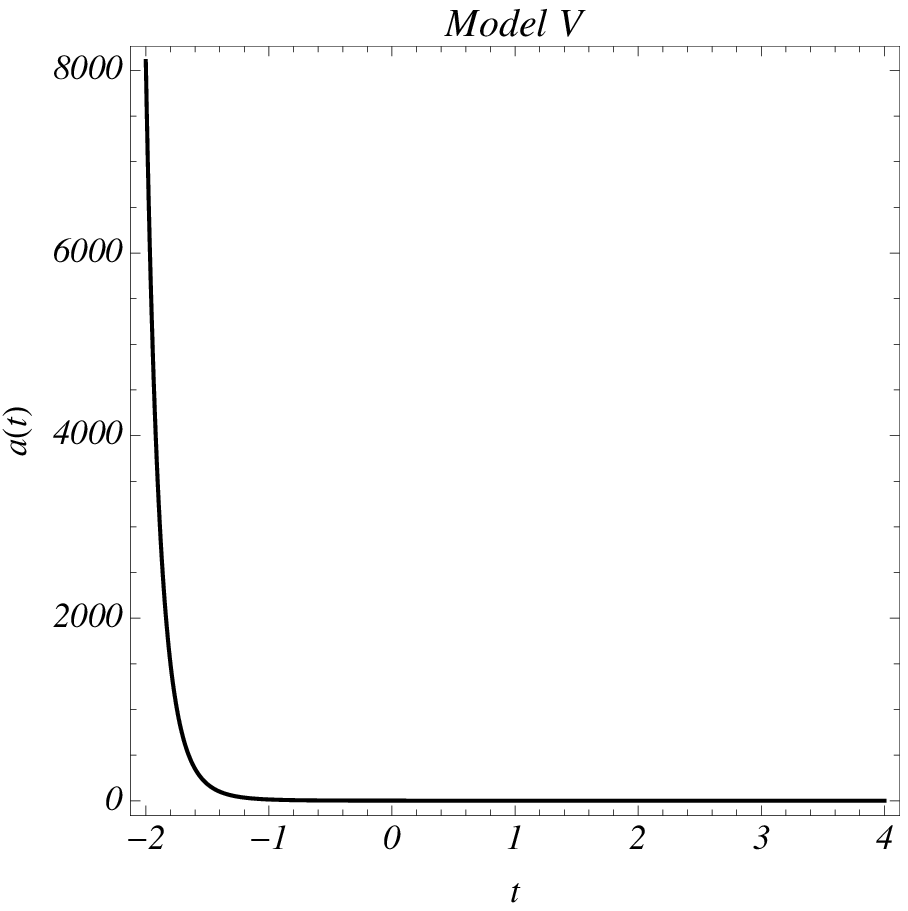}
\includegraphics[scale=0.5]{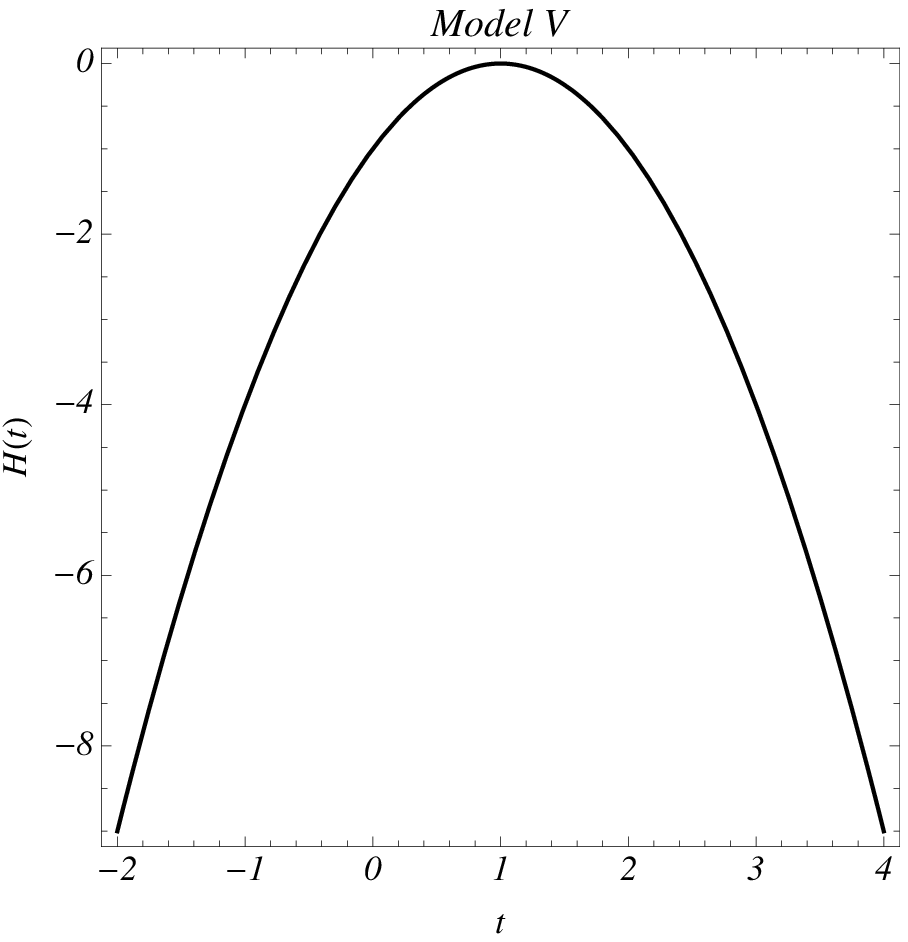}
\includegraphics[scale=0.5]{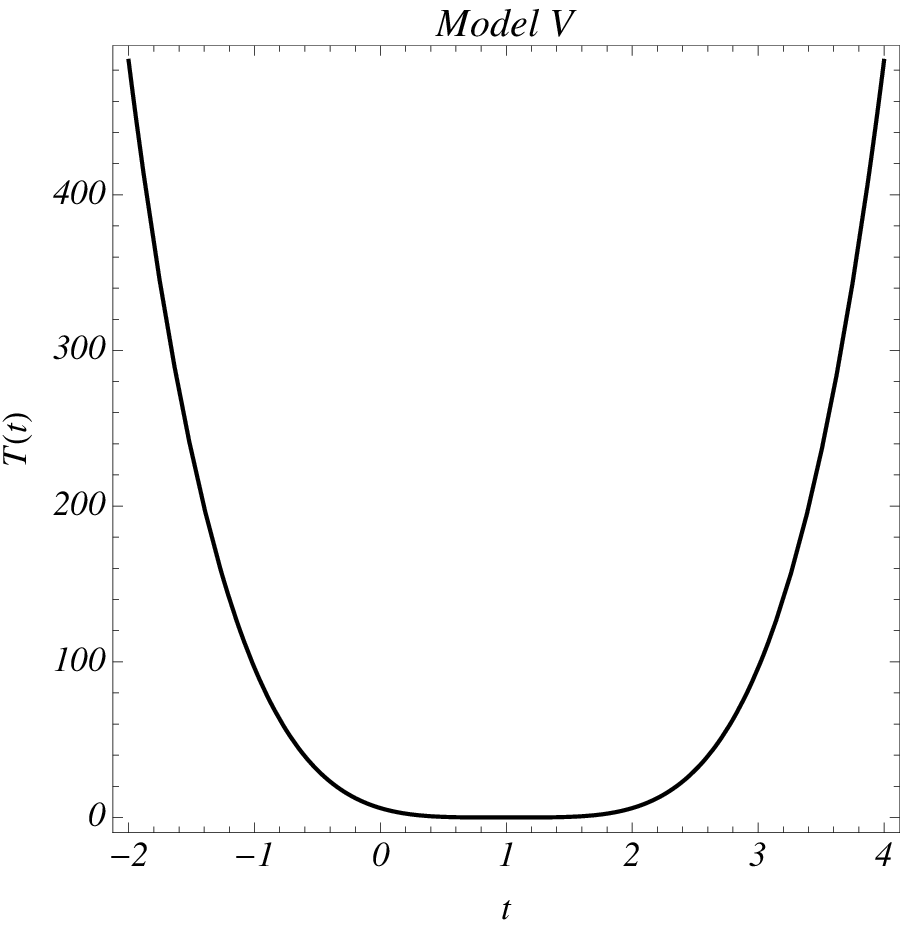}
\caption{A sample for each model analysed in the manuscript, where the evolution of the scale factor, the Hubble parameter and the torsion tensor are depicted for a particular set of the free parameters of the models. The bouncing character of the solutions is clearly shown as well as the possible singularities that may occur.}
\label{Model1}
\end{figure}

Let us start by considering a bouncing solution described by an scale factor with an exponential evolution,
\begin{equation}
a(t) = A \exp \left(\alpha \dfrac{t^2}{{t_*}^2}\right),
\end{equation}
where $t_*$ is some arbitrary time, $A > 0$ and $\alpha > 0$ are constants. By evaluating the expression at $t = 0$, it can easily be concluded that $a(0) = A$. In such cases, $H$ is given by
\begin{equation}
H = \dfrac{2\alpha t}{{t_*}^2}.
\end{equation}
This means that a bounce is located at $t = 0$ since $H < 0$ for $t < 0$, $H = 0$ at $t = 0$ and $H > 0$ for $t > 0$. Consequently, $T$ and $T_G$ are given by
\begin{align}
T &= 6H^2 = \dfrac{24\alpha^2 t^2}{{t_*}^4}, & T_G &= \dfrac{8\alpha}{{t_*}^2} T + \dfrac{2T^2}{3}.
\label{TTG11}
\end{align}
Furthermore, the scale factor can be solely expressed in terms of the torsion scalar $T$ as
\begin{equation}
a(T) = a(0) \exp \left(\dfrac{T{t_*}^2}{24\alpha}\right) = a(0) \exp \left(\alpha\dfrac{T}{T_*}\right),
\end{equation}
where $T_* \equiv T(t=t_*) = 24\alpha^2/{t_*}^2$. \newline

In order to solve the Friedmann equations for this model, some particular ansatzs for the gravitational Lagrangian are considered. Before doing so, let us first simplify the stress-energy component of the field equations by setting $a(t_0) = 1$ at some arbitrary time $t_0 > 0$, such that:
\begin{equation}
{t_0}^2 = -\dfrac{{t_*}^2}{\alpha} \ln A.
\end{equation}
Since $\alpha$ is positive, the equation yields real values for the time $t_0$ if only if $0 < A < 1$. Thus, as long as the value of $A$ is restricted in the range $A \in (0,1)$, one can define the parameters $T_0 \equiv T(t = t_0)$ and $\Omega_{w_i,0} \equiv \Omega_{w_i}(t = t_0)$, which may describe their present time values.
\subsection{$f(T,T_G) = g(T) + h(T_G)$}\label{sec:addition_bouncingI}

By assuming a gravitational Lagrangian of the type $f(T,T_G) = g(T) + h(T_G)$, the Friemdann equation becomes
\begin{equation}
g + h - 2T g_T - T_G h_{T_G} + 24 H^3 h_{T_{G}T_G} \dot{T_G} = T_0 \sum\limits_i \Omega_{w_i,0} a^{-3(1+w_i)}.
\label{model11}
\end{equation}
Since the scale factor can be expressed in terms of $T$ only, the differential equation (\ref{model11}) can be split into a pair of equations as follows
\begin{align}
&g - 2T g_T = T_0 \sum\limits_i \Omega_{w_i,0} A^{-3(1+w_i)} \exp \left[-\dfrac{(1+w_i) T{t_*}^2}{8\alpha}\right], \label{model11a} \\
&h - T_G h_{T_G} + 2 h_{T_{G}T_G} \left({T_G}^2 - \dfrac{4T^4}{9}\right) = 0. \label{model11b}
\end{align} 
Then, the solution for $g$ yields
\begin{equation}
g(T) = c_1 \sqrt{T}+T_0 \sum\limits_i \Omega_{w_i,0} A^{-3 (1+w_i)}\left[\sqrt{\pi} \ x_i \ \text{erf}\left(x_i \right)+{\rm e}^{-{x_i}^2} \right],
\end{equation}
where $c_1$ is a constant of integration (which corresponds to the DGP term) and $x_i \equiv \sqrt{\dfrac{T (1+w_i){t_*}^2}{8 \alpha}}$. \newline

In order to solve the equation (\ref{model11b}) for $h$, we rewrite the equation as follows
\begin{equation}
h(x)-\frac{x \left(x^2+84 \dfrac{\alpha}{{t_*}^2}  x+288 \dfrac{\alpha^2}{{t_*}^4}\right)}{2 \left(x+12 \dfrac{\alpha}{{t_*}^2} \right)^2}h'(x)+\frac{24 \dfrac{\alpha}{{t_*}^2} x^2}{x+12 \dfrac{\alpha}{{t_*}^2}}h''(x) = 0,
\label{eqmodel11b}
\end{equation}
where 
\begin{equation}
x \equiv \sqrt{6T_G+144 \dfrac{\alpha^2}{{t_*}^4}} - 12\dfrac{\alpha}{{t_*}^2}\ .
\end{equation}
Here we have used (\ref{TTG11}). The solution of equation (\ref{eqmodel11b}) yields
\begin{align}
h(x) &= x \left(x+24 \dfrac{\alpha}{{t_*}^2}\right)c_1 + c_2 \exp \left(\dfrac{x{t_*}^2}{48 \alpha}\right) \bigg[-6 \sqrt{\dfrac{\alpha}{{t_*}^2}  x} \left(x+48\dfrac{\alpha}{{t_*}^2}\right) \nonumber \\
&+\sqrt{3} x \left(x+24\dfrac{\alpha}{{t_*}^2}\right) F\left(\dfrac{1}{4}\sqrt{\dfrac{x{t_*}^2}{3\alpha}}\right)\bigg]
\end{align}
where $c_1$ and $c_2$ are constants of integration and $F(z)$ is the Dawson integral which is defined as
\begin{equation}
F(z) \equiv {\rm e}^{-z^2} \int\limits_0^z {\rm e}^{y^2} {\rm d}
y.
\end{equation}
Next step would be to check the existence of vacuum solutions, i.e. $f(0,0) = g(0) + h(0) = 0$. In this case, $h(T_G = x = 0)$ is equal to 0. Thus, we require $g(T = 0) = 0$. However, the resulting limit is
\begin{equation}
g(0) = T_0 \sum\limits_i \Omega_{w_i,0} A^{-3 (1+w_i)},
\end{equation}
which is trivially satisfied in vacuum, where $\Omega_{w_i,0}=0$.
\subsection{$f(T,T_G) = T g(T_G)$}\label{sec:bouncingI-model2}

When considering a $T$ rescaling-type models, the Friedmann equations become:
\begin{align}
&g - T_G g_{T_G} + \dfrac{4T^2}{3} g_{T_G} - 2 g_{T_{G}T_G} \left({T_G}^2 - \dfrac{4T^4}{9}\right) \nonumber \\
&= -\dfrac{T_0}{T} \sum\limits_i \Omega_{w_i,0} A^{-3(1+w_i)} \exp \left[-\dfrac{(1+w_i) T{t_*}^2}{8\alpha}\right].
\end{align} 
Similar to the previous case, the Friedmann equation has to be fully expressed in terms of $T_G$. By using the substitution  $x \equiv \sqrt{6T_G+144 \dfrac{\alpha^2}{{t_*}^4}} - 12\dfrac{\alpha}{{t_*}^2}$, the resulting equation is given by
\begin{align}
&g(x)+\frac{x \left({t_*}^4 x^2+36 \alpha {t_*}^2 x-288 \alpha ^2\right)}{2 \left(12 \alpha +{t_*}^2 x\right)^2}g'(x)-\frac{24 \alpha  x^2}{12 \alpha +{t_*}^2 x}g''(x) \nonumber \\
&=-2T_0 \sum\limits_i \frac{\Omega_{w_i,0} A^{-3(1+w_i)}}{x}\exp\left[-\frac{(1+w_i) x{t_*}^2}{16 \alpha}\right],
\end{align} 
The solution of this equation can be found by a power series, such that:
\begin{equation}
g(x) = \sum\limits_{n = 0}^\infty a_n x^{n+r}+g_{part}(x)\ ,
\end{equation}
where $g_{part}(x)$ corresponds to the particular solution of the inhomogeneous equation, while the exponent $r = 1$ and $r = -1/2$ for the homogeneous equation with the recurrence relation:
\begin{align}
(n+r){t_*}^4 a_{n-2} + 12\left[4+ (n+r-1)(11-4n-4r)\right]{t_*}^2 \alpha a_{n-1} 
-288(n+r-1)(2 n+2 r+1)\alpha^2 a_n = 0,
\end{align}
with $a_{-2} = a_{-1} = 0$, which yields the following solutions for the homogeneous part of the equation:
 \begin{align}
g_1(x) &= \sum\limits_{n = 0}^\infty a_n x^{n+1} = \frac{7 c_1}{48 {t_*}^5} \bigg[12 \left({t_*}^5 x^2+84 \alpha {t_*}^3 x-864 \alpha^2 {t_*}\right)\nonumber \\
&+\sqrt{\frac{3 \pi }{\alpha x}} \left({t_*}^6 x^3+108 \alpha {t_*}^4 x^2+20736 \alpha ^3\right) \text{erf}\left(\sqrt{\frac{{t_*}^2 x}{48 \alpha}} \right)\exp \left(\dfrac{{t_*}^2 x}{48 \alpha }\right)\bigg], \\
g_2(x) &= \sum\limits_{n = 0}^\infty a_n x^{n-\frac{1}{2}} = \dfrac{{t_*}^6 x^3+108 \alpha {t_*}^4 x^2+20736 \alpha ^3}{216 \alpha ^2 {t_*}^2 \sqrt{x}}\exp \left[\dfrac{{t_*}^2 x}{48 \alpha }\right] c_1 ,
\end{align}
where $\text{erf}(z)=\frac{2}{\sqrt{\pi }}\int _0^z {\rm e}^{-t^2} {\rm d}
t$ is the error function. Both solutions satisfy the vacuum constraint $g_i(0)=0$. Finally, the particular solution can be found by using a Green function $G(x,s) = \dfrac{g_1(s)g_2(x) - g_2(s)g_1(x)}{W(s)}$, where $W(s)$ holds for the Wronskian. Nevertheless, an analytical solution is not possible to be found. However, the resulting particular solution at $T \rightarrow 0$ limit (which corresponds to $x \rightarrow 0$) is defined since the integral would be equal to zero and since both $g_1(0) = g_2(0) = 0$,  would imply  $g_{part}(0) = 0$.

\subsection{$f(T,T_G) = T_G g(T)$}

For a similar type of models where a rescaling of $T_G$ is included, the Friedmann equation becomes
\begin{equation}
-\dfrac{4T^3}{3} g_{T} = T_0 \sum\limits_i \Omega_{w_i,0} A^{-3(1+w_i)} \exp \left[-\dfrac{(1+w_i) T{t_*}^2}{8\alpha}\right],
\end{equation}
whose solution is given by
\begin{align}
g(T) &= c_1 +\sum\limits_i \dfrac{3\Omega_{w_i,0} T_0 A^{-3(1+w_i)}}{8T^2} \bigg\lbrace \left[1-\frac{T (1+w_i){t_*}^2}{8 \alpha }\right]\exp \left[-\frac{T (1+w_i){t_*}^2}{8 \alpha}\right]\nonumber \\
&-\left[\dfrac{T(1+w_i){t_*}^2}{8\alpha}\right]^2 \text{Ei}\left[-\frac{T (1+w_i){t_*}^2}{8 \alpha}\right]\bigg\rbrace,
\end{align}
where $c_1$ is an integration constant and Ei$(z)$ is the exponential integral function $\text{Ei}(z) \equiv -\int\limits_{-z}^{\infty} \dfrac{{\rm e}^{-y}}{y} {\rm d}y.$
The solution can be expressed in a more compact form by making use of the substitution variable $x_i \equiv -\dfrac{T (1+w_i){t_*}^2}{8 \alpha}$, which results into
\begin{align}
g(x) &= c_1 +\sum\limits_i \dfrac{3 T_0 \Omega_{w_i,0} A^{-3(1+w_i)} (1+w_i)^2{t_*}^4}{512\alpha^2} \left[\dfrac{1+x_i}{{x_i}^2}{\rm e}^{x_i} - \text{Ei}(x_i)\right].
\end{align} 
For vacuum solutions, we require the Lagrangian $f(T = x = 0) = 0$. In this case, we find
\begin{equation}
f(0) = -\frac{T_0}{8}\sum\limits_i \Omega_{w_i,0} A^{-3 (1+w_i)}\left[1+3w_i+3 (1+w_i) \dfrac{{\rm e}^{x_i}}{x_i}\right]\bigg|_{x_i \rightarrow 0}.
\end{equation}
In the case of vacuum, the condition is satisfied although the resulting Lagrangian would only be composed of the Gauss-Bonnet term, which effectively does not contribute to the field equations.  On the other hand, for single fluids, the Lagrangian diverges even for the case $w_i = -1$. For a cosmological constant like fluid, $x_i$ is already 0 by definition hence requiring more attention when taking the limit. The Lagrangian as $T \rightarrow 0$ in the presence of this fluid becomes
\begin{equation}
f(0) = \frac{\Omega_{-1,0} T_0}{4}+\frac{3 \alpha  \Omega_{-1,0} T_0}{T{t_*}^2}\bigg|_{T \rightarrow 0},
\end{equation}
which diverges as $T \rightarrow 0$. 

\subsection{$f(T,T_G) = -T + T_G g(T)$}

Whether we consider a model expressed as a correction to the Teleparallel action with a rescaling of $T_G$, the Friedmann equation becomes
\begin{equation}
T- \dfrac{4T^3}{3} g_T = T_0 \sum\limits_i \Omega_{w_i,0} A^{-3(1+w_i)} \exp \left[-\dfrac{(1+w_i) T{t_*}^2}{8\alpha}\right],
\end{equation} 
whose solution is identical to the previous model with an extra particular solution of the form
\begin{equation}
g_{part}(T) = -\dfrac{3}{4T}.
\end{equation}
The contribution to the Lagrangian in this case is given by
\begin{equation}
f_{part}(T,T_G) = -\dfrac{3T_G}{4T} = -\dfrac{6\alpha}{{t_*}^2} - \dfrac{T}{2},
\end{equation}
which in the $T\rightarrow 0$ limit reduces to a non-zero constant. As in the previous case there is no trivial vacuum solution.


\subsection{$f(T,T_G) = -T + \mu\left(\dfrac{T}{T_0}\right)^\beta\left(\dfrac{T_G}{T_{G,0}}\right)^\gamma$}

For TEGR with a power-law model, the Friedmann equation becomes
\begin{align}
\mu\left(\dfrac{T}{T_0}\right)^{\beta+\gamma}\left(\dfrac{12\alpha + T{t_*}^2}{12\alpha + {T_0}{t_*}^2}\right)^\gamma  \left[1- 2\beta - \gamma + \dfrac{24\alpha\beta\gamma}{12\alpha  + T {t_*}^2} + \dfrac{48\gamma(\gamma-1)\alpha \left(6\alpha+T{t_*}^2\right)}{\left(12\alpha+T{t_*}^2\right)^2} \right] 
+ T = T_0 \sum\limits_i \Omega_{w_i,0} a^{-3(1+w_i)},
\end{align}  
where $\mu$, $\beta$ and $\gamma$ are constants. By evaluating the expression at current times, the value of $\mu$ is found to be
\begin{align}
\mu 
= \dfrac{T_0 \left(-1 + \sum\limits_i \Omega_{w_i,0}\right)}{1- 2\beta - \gamma + \dfrac{24\alpha\beta\gamma}{12\alpha  + T_0 {t_*}^2} + \dfrac{48\gamma(\gamma-1)\alpha \left(6\alpha+T_0{t_*}^2\right)}{\left(12\alpha+T_0{t_*}^2\right)^2}} 
\equiv \dfrac{T_0 \left(-1 + \sum\limits_i \Omega_{w_i,0}\right)}{\nu},
\end{align}  
where $\nu$ is defined as the denominator. The expression is true provided that $\nu \neq 0$. To obtain vacuum solutions, the following condition must be satisfied
\begin{equation}
\beta + \gamma > 0.
\end{equation}
Using this, since when $t = 0$, $T = 0$, another condition has to be obeyed
\begin{align}
&\sum\limits_i \Omega_{w_i,0}  A^{-3(1+w_i)} = 0.
\end{align} 
This condition is satisfied in the case of vacuum. However, this condition cannot be satisfied in the existence of fluids since $\Omega_{w_i,0}, A > 0$. Therefore, we only consider the former. From the definition of $\mu$, the Friedmann equation simplifies to
\begin{align}
&\left(\dfrac{T}{T_0}\right)^{\beta+\gamma}\left(\dfrac{12\alpha + T{t_*}^2}{12\alpha + {T_0}{t_*}^2}\right)^\gamma  \left[1- 2\beta - \gamma + \dfrac{24\alpha\beta\gamma}{12\alpha  + T {t_*}^2} + \dfrac{48\gamma(\gamma-1)\alpha \left(6\alpha+T{t_*}^2\right)}{\left(12\alpha+T{t_*}^2\right)^2} \right] \nonumber \\
&= \nu\dfrac{T}{T_0},
\end{align} 
In order to determine which values of $\beta$ and $\gamma$ satisfy this equation, the equation must hold at all times. The equation trivially holds when $t = 0$, however this must also hold for arbitrary time. Thus, the time dependent (torsion scalar) terms must cancel. The only possible solution is $\gamma = 0, \beta = 1$, which sets $\nu = -1$ and consequently $\mu = T_0$. However, this implies that $f(T,T_G) = 0$ which is not physical. Thus, a power-law solution with a TEGR contribution cannot describe this bouncing cosmology.

\section{Bouncing model II: Oscillatory model}
\label{model2}

The second bouncing model we are considering here is described by an oscillatory scale factor:
\begin{equation}
a(t) = A \sin^2 \left(B \dfrac{t}{t_*}\right),
\end{equation}
where $t_* > 0$ is some reference time, $A > 0$ and $B > 0$ are dimensionless constants. Here, the restrictions for $t_*$ and $B$ can be relaxed to simply be non-zero. The choice here helps defining the subsequent parameters and ease the analysis for determining which models obey the necessary conditions. For such model, the Hubble parameter is
\begin{equation}
H = \dfrac{2B}{t_*} \cot \left(B \dfrac{t}{t_*}\right).
\end{equation}
This oscillatory model produces two different bounces. For times $t = \frac{n\pi t_*}{B}$, $n \in \mathbb{Z}$, the model describes the time when the universe reaches a crunch ($a = 0, H \rightarrow -\infty$) and rebirths with a big bang ($a = 0, H \rightarrow \infty$). This corresponds to a superbounce. On the other hand, for times $t = \frac{(2n+1)\pi t_*}{2B}$, $n \in \mathbb{Z}$, the universe reaches maximum size with no further expansion ($a = A, H = 0$). This also corresponds to a bounce since $H$ transitions from positive, zero to negative, before, at and after the maximum peak. In this case, $T$ and $T_G$ are 
\begin{align}\label{eq:bouncingIII-T-TGrel}
T &= 6H^2 = \dfrac{24B^2}{{t_*}^2} \cot^2 \left(B \dfrac{t}{t_*}\right), & T_G &= 4T\left(\dfrac{T}{12}-\frac{2 B^2}{{t_*}^2}\right).
\end{align}
Using these definitions, the scale factor can be expressed in terms of the torsion scalar to be
\begin{equation}
a(T) = \dfrac{A}{1+\dfrac{T{t_*}^2}{24B^2}}.
\end{equation}
Before solving for the gravitational actions considered above, we first assume the existence of some time $t_0 > 0$ at which the scale factor is 1,
\begin{equation}
\dfrac{1}{A} = \sin^2 \left(B \dfrac{t_0}{t_*}\right).
\end{equation}
Since $0 \leq \sin^2(x) \leq 1$, $A > 1$ is required. If we set our first big bang to be at $t = 0$ for instance and the first maximum of the expansion at $t = \frac{\pi t_*}{2B}$, then the present time would lie at $t_0 = \frac{t_* \sin ^{-1}\left(\frac{1}{A}\right)}{B}$. With this time defined, the remaining present-time parameters, such as the $\Omega_{w_i,0}$  density parameters, current times torsion scalar $T_0 \equiv T(t = t_0) = \frac{24 \left(A^2-1\right) B^2}{{t_*}^2}$ and so on, can be defined. 

\subsection{$f(T,T_G) = g(T) + h(T_G)$}

For type of model, the Friedmann equation results
\begin{align}
&g + h - 2T g_T - T_G h_{T_G} -\frac{4 T \left(2 T^2-3 T_G\right) \left(T {t_*}^2-12 B^2\right)}{9 {t_*}^2} h_{T_{G}T_G} \nonumber \\
&= T_0 \sum\limits_i \Omega_{w_i,0} A^{-3(1+w_i)}\left(1+\dfrac{T{t_*}^2}{24B^2}\right)^{3(1+w_i)}.
\end{align} 
Before solving the ODE, we point out that since the $T$ and $T_G$ are related through a quadratic expression (\ref{eq:bouncingIII-T-TGrel}), the torsion scalar can be expressed in terms of $T_G$ as
\begin{equation}\label{eq:bouncingIII-T-TGrel_2}
T = \frac{12B^2}{{t_*}^2} \left(1 - \sqrt{1+\frac{T_G {t_*}^4}{48 B^4}}\right),
\end{equation}
where the plus solution is neglected since it is inconsistent at maximum size periods (i.e. when $T = T_G = 0$). In doing so, the ODE can be separated into two ODEs, for $g$ and for $h$. This is only possible provided their respective ODEs result into a constant. It turns out that similar to other bouncing models, this constant drops out of the Lagrangian so it is neglected from the solutions. The resulting ODEs to solve are the following
\begin{align}
&g - 2T g_T = T_0 \sum\limits_i \Omega_{w_i,0} A^{-3(1+w_i)}\left(1+\dfrac{T{t_*}^2}{24B^2}\right)^{3(1+w_i)}, \\
&h - T_G h_{T_G} -\frac{4 T \left(2 T^2-3 T_G\right) \left(T {t_*}^2-12 B^2\right)}{9 {t_*}^2} h_{T_{G}T_G} = 0,
\end{align}
where $T$ is expressed in terms of $T_G$ in the ODE for $h$. 


The solution for $g(T)$ is given by:
\begin{align}
&g(T) = c_1 \sqrt{T}-T_0 \sum\limits_i \frac{\Omega_{w_i,0} A^{-3 (1+w_i)}}{69120 B^6} \bigg\lbrace T^3 {t_*}^6 \, _2F_1\left(\frac{5}{2},-3 w_i;\frac{7}{2};-\frac{T {t_*}^2}{24 B^2}\right) \nonumber \\
&+120 B^2 T {t_*}^2 \left[72 B^2 \, _2F_1\left(\frac{1}{2},-3 w_i;\frac{3}{2};-\frac{T {t_*}^2}{24 B^2}\right)+T {t_*}^2 \, _2F_1\left(\frac{3}{2},-3 w_i;\frac{5}{2};-\frac{T {t_*}^2}{24 B^2}\right)\right]\nonumber \\
&-69120 B^6 \, _2F_1\left(-\frac{1}{2},-3 w_i;\frac{1}{2};-\frac{T {t_*}^2}{24 B^2}\right)\bigg\rbrace,
\end{align}
where $c_1$ is an integration constant corresponding to the DGP contribution in the Lagrangian. 
While the solution for $h(T_G)$ equation leads to
\begin{equation}
h(x) = x(x-2)c_1 + \left[2 (8-3 x) \sqrt{x}-3 \sqrt{2} x(x-2) \tan ^{-1}\left(\sqrt{\frac{x}{2}}\right)\right]c_2,
\end{equation}
where $c_{1,2}$ are integration constants and $x \equiv 1 - \sqrt{1+\dfrac{T_G {t_*}^4}{48 B^4}}$. In the vacuum limit $T_G \rightarrow 0$ (or equivalently, $x \rightarrow 0$), $h(0) = 0$ whereas in the limit $T \rightarrow 0$, the resulting function leads to $g(0) = T_0\sum\limits_i \Omega_{w_i,0} A^{-3 (1+w_i)}$. Hence,  $f(0,0)=0$ is only possible in absence of fluids, $\Omega_{w_i,0}=0$, as natural.


%

\subsection{$f(T,T_G) = T g(T_G)$}

For a TEGR rescaling model, the Friedmann equation is given by
\begin{align}
&g + \left(\dfrac{4T^2}{3}- T_G\right) g_{T_G} +\frac{4 T \left(2 T^2-3 T_G\right) \left(T {t_*}^2-12 B^2\right)}{9 {t_*}^2} g_{T_G T_G} \nonumber \\
&= -\dfrac{T_0}{T} \sum\limits_i \Omega_{w_i,0} A^{-3(1+w_i)}\left(1+\dfrac{T{t_*}^2}{24B^2}\right)^{3(1+w_i)}.
\end{align} 
Using Eq. \ref{eq:bouncingIII-T-TGrel_2}, the ODE can be expressed fully in terms of $T_G$ and hence can be solved for $g$. To simplify the ODE, we make a change of variables by introducing the variable $x \equiv 1 - \sqrt{1+\dfrac{T_G {t_*}^4}{48 B^4}}$. This results into
\begin{align}
&g(x)+\frac{x (x^2-5x-2)}{2 (x-1)^2}g'(x)+\frac{x^2(x+2)}{x-1}g''(x) \nonumber \\
&= -T_0 \sum\limits_i \Omega_{w_i,0} \frac{{t_*}^2 2^{-3w_i-5}}{3B^2 x}\left(\frac{A}{x+2}\right)^{-3(1+w_i)}.
\end{align}
%
%
The homogeneous solution can be expressed by a power-series leading to
\begin{equation}
g(x) = \sum\limits_{n=0}^\infty a_n x^n,
\end{equation}
where the following recurrence relation is obtained
\begin{equation}
2 (n-1)^2 a_n+n a_{n-2}+(n-5) (2 n-1) a_{n-1}=4 n (n+1) a_{n+1}, 
\end{equation}
with $a_{-2} = a_{-1} = 0$. A general solution to the recurrence relation can not be found. Nonetheless, the first few terms of the series are found to be
\begin{align}
a_0 = 0, a_1 = 1, a_2 = 0, a_3 = -\frac{3}{8}a_1,a_4 = 0, a_5 = \frac{21}{640}a_1, a_6 = -\frac{11}{1600}a_1.
\end{align}
Thus, the first solution of the homogeneous equation is 
\begin{equation}
g_1(x) = \sum\limits_{n=0}^\infty a_n x^n = a_1 \left(x-\frac{3 x^3}{8}+\frac{21 x^5}{640}-\frac{11 x^6}{1600}+\dots\right),
\end{equation}
where $a_1$ takes the role of the integration constant. In order to find the second solution, one can use Abel's identity although only in certain intervals \cite{ross2013differential}. By using Abel's identity, this results into
\begin{equation}
g_2(x) = C g_1(x) \int^x \dfrac{(1-\eta)}{\sqrt{\eta}(2+\eta) {g_1}^2(\eta)} \; {\rm d}\eta.
\end{equation}
where $C$ is an integration constant. However, the above homogeneous solution is only applicable for $x \in (0, 1) \cup (1, \infty)$, as $g_1(x)$ and its derivative are not continuous at $x=0$ and $x=1$ respectively. Finally, since the power-series is not expressed in terms of some analytical function, integrating over an infinite series is intractable. We also point out that in the vacuum limit, $g_1(0) = 0$ although nothing can be inferred about $g_2(0)$. 


\subsection{$f(T,T_G) = T_G g(T)$}

For a TEGB rescaling, the Friedmann equation is given by
\begin{equation}
- \dfrac{4T^3}{3}g_T  = T_0 \sum\limits_i \Omega_{w_i,0} A^{-3(1+w_i)}\left(1+\dfrac{T{t_*}^2}{24B^2}\right)^{3(1+w_i)}.
\end{equation} 
To simplify this equation, we introduce a change of variables defined by $x \equiv 1+\dfrac{T{t_*}^2}{24B^2}$, such that it leads:
\begin{equation}
(x-1)^3 g_x  = \sum\limits_i \xi_{w_i} x^{3(1+w_i)},
\end{equation} 
where $\xi_{w_i} \equiv -\frac{3}{4}\left[\frac{{t_*}^2}{24 B^2}\right]^2 T_0 \Omega_{w_i,0} A^{-3(1+w_i)}$. Depending on the value of $w_i$, we have different particular solutions. Due to the sum being finite, the sum of the particular solutions corresponding to each $w_i$ will be the general solution. \\

\paragraph{Case 1: $w \neq n/3$, $n \in \mathbb{Z}, n \geq -1$}

 For this set of values, the solution is given by
\begin{align}
g(x) &= \frac{\xi_{w_i}}{2}  x^{1+3 w_i} \bigg\lbrace -\frac{1}{3 w_i (x-1)^3}\bigg[(x-1) (2-x+3 w_i x) \, _2F_1\left(1,1;1-3 w_i;\frac{1}{1-x}\right)\nonumber \\
&+x (4+3 w_i-5x-3 w_i x) \, _2F_1\left(1,2;1-3 w;\frac{1}{1-x}\right)\bigg]+\frac{2}{1+3w_i}\bigg\rbrace.
\end{align}
For every $w_i$, the Lagrangian diverges in the vacuum limit. \\

\paragraph{Case 2: $w = n/3$, $n \in \mathbb{Z}, n \geq -1$} 

For the remaining set of values, we solve the ODE as follows:
\begin{equation}
(x-1)^3 g_x  = \xi_{w_i} x^{3+n},
\end{equation}
where the summation is suppressed for simplicity. Next, we define the variable $y \equiv x-1$ to transform the ODE into
\begin{equation}
g_y = \xi_{w_i} \dfrac{(y+1)^{3+n}}{y^3}.
\end{equation}
Since $n \in \mathbb{Z}, n \geq -1$, by the Binomial theorem, the binomial term can be expanded as
\begin{equation}
(y+1)^{3+n} = \sum\limits_{k=0}^{3+n} \binom{n}{k} y^k.
\end{equation}
Therefore, the resulting solution is given by
\begin{equation}
g(y) = \xi_{w_i} \sum\limits_{k=0}^{3+n} \binom{n}{k} \int y^{k-3} \; {\rm d}y.
\end{equation}
For these values, the Lagrangian diverges in the vacuum limit.

%

\subsection{$f(T,T_G) = -T + T_G g(T)$}

For a TEGB rescaling with TEGR, the Friedmann equation is given by
\begin{equation}
T - \dfrac{4T^3}{3} g_T = T_0 \sum\limits_i \Omega_{w_i,0} a^{-3(1+w_i)},
\end{equation} 
whose solution is
\begin{equation}
g(T) = c_1 - \dfrac{3}{4T} + h(T),
\end{equation}
where $c_1$ is a constant of integration corresponding to the Gauss-Bonnet contribution in the Lagrangian and $h(T)$ is the solution found in the previous model. In the vacuum limit, the Lagrangian is
\begin{equation}
f(0,0) = \frac{6 B^2}{{t_*}^2} + T_G h(T)|_{T,T_G \rightarrow 0}.
\end{equation} 
Following the discussions in the previous section, the last term is finite only in vacuum leading to $T_G h(T)|_{T,T_G \rightarrow 0} = 0$. However, since $B, t_* > 0$, the Lagrangian does not satisfy the vacuum condition. Therefore, this model cannot describe the oscillating cosmology whilst obeying the vacuum condition.

\subsection{$f(T,T_G) = -T + \mu\left(\dfrac{T}{T_0}\right)^\beta\left(\dfrac{T_G}{T_{G,0}}\right)^\gamma$}

For a power-law model with a TEGR contribution, the Friedmann equation becomes
\begin{align}
&T + \mu\left(\dfrac{T}{T_0}\right)^\beta\left(\dfrac{T_G}{T_{G,0}}\right)^\gamma  \bigg[1- 2\beta - \gamma + \beta\gamma\left(2-\dfrac{4T^2}{3T_G}\right) \nonumber \\
&- \dfrac{\gamma(\gamma-1)}{9 {t_*}^2}\frac{4 T \left(2 T^2-3 T_G\right) \left(T {t_*}^2-12 B^2\right)}{{T_G}^2}\bigg] = T_0 \sum\limits_i \Omega_{w_i,0} a^{-3(1+w_i)}.
\end{align} 
The Lagrangian satisfies the vacuum condition as long as $\beta + \gamma > 0$. At times when $T = T_G = 0$ (which occurs at the maximum universe size), the Friedmann equation yields the following condition,
\begin{align}
&0 = \sum\limits_i \Omega_{w_i,0} A^{-3(1+w_i)}.
\end{align} 
However, this is possible only in vacuum. Then, the Friedmann equation can be evaluated at current times to evaluate $\mu$,
\begin{align}
&\mu = \dfrac{-T_0}{1- 2\beta - \gamma + \beta\gamma\left(2-\dfrac{4{T_0}^2}{3{T_{G,0}}}\right) - \dfrac{\gamma(\gamma-1)}{9 {t_*}^2}\frac{4 T_0 \left(2 {T_0}^2-3 T_{G,0}\right) \left(T {t_*}^2-12 B^2\right)}{{T_{G,0}}^2}} \nonumber \\
&\equiv -\dfrac{T_0}{\nu},
\end{align} 
where $\nu \neq 0$ is defined as the denominator. This simplifies the Friedmann equation into
\begin{align}
&T -\dfrac{T_0}{\nu}\left(\dfrac{T}{T_0}\right)^\beta\left(\dfrac{T_G}{T_{G,0}}\right)^\gamma  \bigg[1- 2\beta - \gamma + \beta\gamma\left(2-\dfrac{4T^2}{3T_G}\right) \nonumber \\
&- \dfrac{\gamma(\gamma-1)}{9 {t_*}^2}\frac{4 T \left(2 T^2-3 T_G\right) \left(T {t_*}^2-12 B^2\right)}{{T_G}^2}\bigg] = 0.
\end{align} 
This equation has to be satisfied at all times. Trivially, this is satisfied when $T = T_G = 0$ and at $t = t_0$, so other time instances are assumed. This allows for a re-arranging of the equation to
\begin{align}
&\left(\dfrac{T}{T_0}\right)^{\beta-1}\left(\dfrac{T_G}{T_{G,0}}\right)^\gamma  \bigg[1- 2\beta - \gamma + \beta\gamma\left(2-\dfrac{4T^2}{3T_G}\right) \nonumber \\
&- \dfrac{\gamma(\gamma-1)}{9 {t_*}^2}\frac{4 T \left(2 T^2-3 T_G\right) \left(T {t_*}^2-12 B^2\right)}{{T_G}^2}\bigg] = \nu.
\end{align} 
Since $\nu$ is a constant, all time dependent (or, equivalently, the torsional and TEGB terms) must vanish. This is possible for the following cases, $\beta = -1, \gamma = 1$ and $\beta = 1, \gamma = 0$. In the former case, although it leads to a non-trivial Lagrangian, the vacuum condition is not satisfied. On the other hand, the latter is the TEGR result which leads to $\nu = -1$ and consequently a zero Lagrangian which is non-physical. Thus, there is no Lagrangian which describes the oscillating cosmology whilst obeying the vacuum condition. \newline

If a Lagrangian composed of the TEGR term with DGP and Gauss-Bonnet terms, the resulting Friedmann equation is given by
\begin{align}
&T = T_0 \sum\limits_i \Omega_{w_i,0} a^{-3(1+w_i)}.
\end{align} 
However, evaluating at times when the universe size is maximum (i.e. $T = T_G = 0)$, yields the previous restriction on the omega parameters
\begin{align}
&0 = \sum\limits_i \Omega_{w_i,0} A^{-3(1+w_i)},
\end{align} 
which is only possible in vacuum. If this is assumed, this sets $T = 0$ at all times which is clearly not the case. Thus, this Lagrangian composition cannot not describe the oscillating cosmology.

\section{Bouncing model III: Power-law model }
\label{model3}

For this section, we consider a scale factor of the form
\begin{equation}
a(t) = \left(\dfrac{t_s - t}{t_0}\right)^{2/c^2},
\end{equation}
where $t_s$ represents the time at which the bounce occurs, $t_0 > 0$ is an arbitrary time parameter which defines the scale factor to be 1 when $t = t_s + t_0$ and $c$ is a constant. In this case, we have the following expressions
\begin{align}
H &= -\dfrac{2}{c^2}\dfrac{1}{t_s-t}, & T &= 6H^2, & T_G &= \dfrac{2T^2}{3}\left(1-\dfrac{c^2}{2}\right).
\end{align}
Furthermore, the scale factor can be solely expressed in terms of the torsion scalar as
\begin{equation}
a(T) = \left(\dfrac{24}{T c^4 {t_0}^2}\right)^{1/c^2}.
\end{equation}
Before continuing further, we make note of the following. We define the following quantities $t_* \equiv t - t_s$ and $\alpha \equiv 2/c^2$. Thus, the scale factor becomes $a(t_*) = \left(t_*/t_0\right)^{\alpha}$, whilst the Hubble parameter, torsion and teleparallel Gauss-Bonnet quantities become
\begin{align}
H &= \dfrac{\alpha}{t_*}, & T &= 6H^2 = 6\dfrac{\alpha^2}{{t_*}^2}, & T_G &= \dfrac{2T^2}{3}\left(1-\dfrac{1}{\alpha}\right).
\end{align}
Note that at $t_* = t_0$, $T_0 \equiv T(t_* = t_0) = 6\alpha^2/{t_0}^2$. This simplifies the expression for the scale factor to be
\begin{equation}
a(T) = \left(\dfrac{T_0}{T}\right)^{\alpha/2}.
\end{equation}
Through this transformation, it effectively simplifies the model to a standard power-law model encountered in single fluid dominated universes with the difference being that multiple fluids are considered. In fact, the Friedmann equation remains unchanged since the time dependent differentiations remain unchanged, being $\dot{T} \equiv dT/dt = dT/dt_*$ and $\dot{T}_G \equiv dT_G/dt = dT_G/dt_*$. Hence, the resulting Friedmann equation is 
\begin{equation}
f - 2T f_T - T_G f_{T_G} -\dfrac{4T^3}{3\alpha} f_{TT_G} - \dfrac{8T^2 T_G}{3\alpha} f_{T_{G}T_G} = T_0 \sum\limits_i \Omega_{w_i,0} \left(\dfrac{T_0}{T}\right)^{-3(1+w_i)\alpha/2}.
\end{equation} 
Let us now find the corresponding Lagrangians for this type of cosmology.

\subsection{$f(T,T_G) = g(T) + h(T_G)$}\label{sec:addition-bouncingIV}

For an additive type model, with two functions $g$ and $h$ of the torsion scalar and TEGB term respectively, the Friedmann equation simplifies to
\begin{equation}
g + h - 2T g_T - T_G h_{T_G} - \dfrac{8T^2}{3\alpha}T_G h_{T_G T_G} = T_0 \sum\limits_i \Omega_{w_i,0} \left(\dfrac{T_0}{T}\right)^{-3(1+w_i)\alpha/2},
\end{equation} 
Note that when $\alpha = 1$ sets $T_G = 0$, one has to be careful in solving the Friedmann equation in this scenario. Thus, we solve the Friedmann equation for the cases when $\alpha = 1$ and $\alpha \neq 1$ separately.\\

For $\alpha = 1$, the function $h$ results into a constant, say $h(T_G) = h(0) = \mu$.\footnote{Note that in principle, $h(0)$ could be divergent. However, in order to satisfy the vacuum condition, this would require that $g(0)$ also divergences and would need to cancel exactly. Thus, for simplicity we shall consider only the finite case.} However, nothing can be inferred on the behaviour of its derivatives, becoming degeneracy. However, we can analyse  the case when the derivatives are constant, i.e. $h'(0) = \beta$ and $h''(0) = \gamma$ for some constants $\beta$ and $\gamma$.  Here, the resulting Friedmann equation is
\begin{equation}
g - 2T g_T = - \mu + T_0 \sum\limits_i \Omega_{w_i,0} \left(\dfrac{T_0}{T}\right)^{-3(1+w_i)\alpha/2},
\end{equation}
whose solution is given by
\begin{equation}
g(T) = c_1 \sqrt{T}-\mu+B_j T\sqrt{\dfrac{T_0}{T}}\ln\left(\dfrac{T_0}{T}\right)+\sum\limits_i A_i \left(\frac{T_0}{T}\right)^{-3\alpha(1+w_i)/2}, 
\end{equation}
for some integration constant $c_1$, whose term corresponds to the DGP term, $A_i \equiv \frac{\Omega_{w_i,0} T_0}{1-3 \alpha  (1+w_i)}$ having $1-3 \alpha  (1+w_i) \neq 0 \; \forall i$ and $B_j \equiv \frac{\Omega_{w_j,0}}{2}$ obeying $1-3 \alpha  (1+w_j) = 0 \; \exists j$. Next, we demand the vacuum condition $f(0,0) = g(0) + h(0) = 0$. This can be satisfied for various scenarios, for instance in vacuum ($B_j = A_i = 0 \; \forall i,j$), for a single fluid obeying the $B_j$ condition, for fluids having EoS $w_i > -1 \; \forall i$ and so on. Examples of functions obeying these set of conditions include $h(T_G) = \sum\limits_{n=1}^i \eta_n \exp \left(\xi_n {T_G}^n\right)$ for $i < \infty$ and constants $\eta_n$ and $\xi_n$ and $h(T_G) = \xi + \sum\limits_{n=1}^i \eta_n {T_G}^n$ for $i < \infty$, and $\xi$ and $\eta_n$ are constants. \newline

Lastly, another solution can be obtained for the case when $h'(0) = \beta$ and $h''(0) \rightarrow \infty$ with $T_G h''(T_G)|_{T_G \rightarrow 0} = \gamma$. In this case, the Friedmann equation reduces to
\begin{equation}
g - 2T g_T  = - \mu + \dfrac{8T^2}{3}\gamma + T_0 \sum\limits_i \Omega_{w_i,0} \left(\dfrac{T_0}{T}\right)^{-3(1+w_i)\alpha/2}.
\end{equation} 
The resulting solution is
\begin{equation}
g(T) =c_1 \sqrt{T}-\frac{8 \gamma  T^2}{9}-\mu+B_j T\sqrt{\dfrac{T_0}{T}}\ln\left(\dfrac{T_0}{T}\right)+\sum\limits_i A_i \left(\frac{T_0}{T}\right)^{-3\alpha(1+w_i)/2},
\end{equation}
where $c_1$, $A_i$ and $B_j$ have the same definitions and conditions as the previous case. The only difference lies in the extra contribution of $-8\gamma T^2/9$ in the Lagrangian. Since in the $T \rightarrow 0$ limit this reduces to 0, the same vacuum conditions obtained previously can be applied. An example of a function with these properties is the function $h(T_G)$ such that $h''(T_G) = \sin(\alpha/T_G)$, for some constant $\alpha > 0$. \newline

In principle, other solutions can be obtained under different conditions, say $h'(0) \rightarrow \infty$ with $T_G h'(T_G)|_{T_G \rightarrow 0} =\beta$ and $h''(0) \rightarrow \infty$ with $T_G h''(T_G)|_{T_G \rightarrow 0} = \gamma$. However, since functions obeying these properties have not been found, these were not considered in the analysis. \\

For $\alpha \neq 1$, the Friedmann equation can be expressed fully in terms of $T$ and $T_G$ as follows
\begin{equation}
g +h - 2T g_T - T_G h_{T_G} - \dfrac{4{T_G}^2}{\alpha-1} h_{T_{G}T_G} = T_0 \sum\limits_i \Omega_{w_i,0} \left(\dfrac{T_0}{T}\right)^{-3(1+w_i)\alpha/2}.
\end{equation} 
which can be split in the following system of equations
\begin{align}
&g - 2T g_T - T_0 \sum\limits_i \Omega_{w_i,0} \left(\dfrac{T_0}{T}\right)^{-3(1+w_i)\alpha/2} = \lambda, \\
&h - T_G h_{T_G} - \dfrac{4{T_G}^2}{\alpha-1} h_{T_{G}T_G} = -\lambda.
\end{align}
Here $\lambda$ is a constant. Hence, the following solutions are obtained,
\begin{align}
&g(T) = \lambda + c_1 \sqrt{T}+B_j T\sqrt{\dfrac{T_0}{T}}\ln\left(\dfrac{T_0}{T}\right)+\sum\limits_i A_i \left(\frac{T_0}{T}\right)^{-3\alpha(1+w_i)/2}, \\
&h(T_G) = -\lambda + T_G c_2 + {T_G}^{\frac{1-\alpha}{4}}c_3, 
\end{align}
where $A_i \equiv \frac{\Omega_{w_i,0} T_0}{1-3 \alpha  (1+w_i)}$ having $1-3 \alpha  (1+w_i) \neq 0 \; \forall i$, $B_j \equiv \frac{\Omega_{w_j,0}}{2}$ obeying $1-3 \alpha  (1+w_j) = 0 \; \exists j$ and $c_{1,2,3}$ are integration constants. The $c_1$ term corresponds to the DGP term while the $c_2$ corresponds to the Gauss-Bonnet term. We also remark that the contribution of $\lambda$ is fictitious since the total contribution of $\lambda$ to the Lagrangian $f$ is zero.

In order to keep vacuum solutions where $g(0) = h(0) = 0$, the following conditions must be satisfied
\begin{align}
\alpha  (1+w_i) > 0 &\implies w_i > -1 \; \forall i, \\
\alpha &< 1.
\end{align}
The first condition is obtained provided that a fluid obeying the $A_i$ condition exists, otherwise the condition is not applicable in vacuum. On the other hand, the second condition holds provided that $c_3 \neq 0$. Otherwise, for cases for which $\alpha \geq 1$, $c_3$ can be set to zero and obtain non-trivial solutions from the $g(T)$ contribution. 

\subsection{$f(T,T_G) = T g(T_G)$}

For a rescaling of $T$ model, the resulting Friedmann equation to solve is
\begin{equation}
g + \left(T_G + \dfrac{4T^2}{3\alpha}\right) g_{T_G} + \dfrac{8T^2}{3\alpha} T_G g_{T_{G}T_G} = -\sum\limits_i \Omega_{w_i,0} \left(\frac{T_0}{T}\right)^{\frac{-3(1+w_i)\alpha+2}{2}}.
\end{equation} 
Similar to the previous case, the equation yields different solutions depending on the values of $\alpha$, i.e. between $\alpha = 1$ and $\alpha \neq 1$.\\

For $\alpha = 1$, $T_G = 0$, such that the function $g(T_G)$ results into a constant, namely $g(T_G) = g(0) = \mu$.\footnote{Similar to the additive case, $h(0)$ could diverge. In spite of this satisfies the vacuum condition, one would require to satisfy the resulting Friedmann equation. Since for this case, we are only interested to illustrate some possible solutions, this case is not considered for simplicity.} Note that for this case, this automatically satisfies the vacuum condition $f(0,0) = 0$. \\

For $\alpha \neq 1$, the Friedmann equation can be expressed fully in terms of $T_G$ as,
\begin{equation}
g + \dfrac{\alpha+1}{\alpha-1} T_G g_{T_G} + \dfrac{4{T_G}^2}{\alpha-1} g_{T_{G}T_G} = -\sum\limits_i \Omega_{w_i,0} \left(\dfrac{T_{G,0}}{T_G}\right)^{\frac{-3(1+w_i)\alpha+2}{4}},
\end{equation} 
which yields a solution of the form
\begin{align}
g(T_G) &= c_1 {T_G}^{m_-}+c_2 {T_G}^{m_+}-\sum\limits_i A_i \left(\frac{T_{G,0}}{T_G}\right)^{\frac{-3(1+w_i)\alpha+2}{4}},
\end{align}
where 
\begin{align}
m_\pm &\equiv \frac{1}{8} \left(3-\alpha \pm \sqrt{\alpha^2 -22\alpha+25}\right), \\
A_i &\equiv \frac{4 (\alpha -1) \Omega_{w_i,0}}{3 \alpha ^2 \left(3 {w_i}^2+7 w_i+4\right)-\alpha  (21 w_i+19)+6},
\end{align}
provided that the denominator of $A_i$ is non-zero $\forall i$, which is satisfied as long as 
\begin{equation}
w_i \neq \frac{7-7 \alpha-\sqrt{\alpha ^2-22 \alpha+25}}{6 \alpha}.
\end{equation}
It is important to distinguish the different solutions stemming from the $c_1$ and $c_2$ contributions. This is done by examining the square root term. The following sub-cases are obtained
\begin{itemize}
\item $\alpha^2-22\alpha+25 > 0$: When the square root is real, this gives the two distinct power-law solutions. Here, the range of values of $\alpha$ obeying the condition are $0 < \alpha < 11-4\sqrt{6}$ and $\alpha > 11+4\sqrt{6}$. In this case, the vacuum condition is satisfied as long as $0 < \alpha < 11-4\sqrt{6}$, otherwise the integration constants are set to zero.
\item $\alpha^2-22\alpha+25 = 0$: In this case, $m_+ = m_-$, effectively combining the two solutions into one $g(T_G) \propto {T_G}^{\frac{3-\alpha}{8}}$. The values of $\alpha$ giving rise to this particular case are $\alpha = 11\pm 4 \sqrt{6}$. In this case, the vacuum condition for this homogeneous solution is satisfied only for $\alpha = 11-4\sqrt{6}$ unless the constant of integration is zero for the other value.
\item $\alpha^2-22\alpha+25 < 0$: When the square root becomes complex, the homogeneous solution has to be re-expressed using the relation
\begin{equation}
a^{b+ic} = a^b \left[\cos(c \ln a) +i \sin(c \ln a)\right].
\end{equation}
For simplicity, we define $i\beta \equiv \sqrt{\alpha^2-22\alpha+25}$. This leads to the following homogeneous solution
\begin{equation}
g_\text{hom.}(T_G) = c_1 {T_G}^{\frac{3-\alpha}{8}}\cos\left(\dfrac{\beta}{8} \ln T_G\right) +c_2 {T_G}^{\frac{3-\alpha}{8}} \sin\left(\dfrac{\beta}{8} \ln T_G\right),
\end{equation}
where the constants of integration $c_1$ and $c_2$ have been redefined. Equivalently, the homogeneous solution can be expressed as
\begin{equation}
g_\text{hom.}(T_G) = c_3 {T_G}^{\frac{3-\alpha}{8}}\cos\left(c_4+\dfrac{\beta}{8} \ln T_G\right),
\end{equation}
where $c_3 \equiv \sqrt{{c_1}^2+{c_2}^2}$ and $c_4 = -\arctan(c_1/c_2)$. In this case, $\alpha$ lies in the range $11-4\sqrt{6} < \alpha < 11+4\sqrt{6}$. For the vacuum condition, we find the following instances. For $11-4\sqrt{6} < \alpha < 7$, the vacuum condition is satisfied whilst for $7 \leq \alpha < 11+4\sqrt{6}$, the latter is satisfied when $c_3 = 0$ i.e. there would be no contribution from the homogeneous solution for this particular range of values.
\end{itemize}
On the other hand, the particular solution satisfies the vacuum condition as long as $w_i > -1, \; \forall i$.

\subsection{$f(T,T_G) = T_G g(T)$}

For this model, the Friedmann equation becomes
\begin{equation}
- \dfrac{4T^3}{3}g_T = T_0 \sum\limits_i \Omega_{w_i,0} \left(\dfrac{T_0}{T}\right)^{-3(1+w_i)\alpha/2},
\end{equation} 
whose solution is given to be
\begin{equation}
g(T) = c_1 + B_j \ln \left(\dfrac{T_0}{T}\right) + \sum\limits_i \dfrac{A_i}{T^2}\left(\frac{T_0}{T}\right)^{-3\alpha (1+w_i)/2},
\end{equation}
where $A_i \equiv \frac{3 \Omega_{w_i,0} T_0}{2 [4-3 \alpha  (1+w_i)]}$ having $4-3 \alpha  (1+w_i) \neq 0 \; \forall i$, $B_j \equiv \frac{3 \Omega_{w_j,0}}{4 T_0}$ obeying $4-3 \alpha  (1+w_j) = 0 \; \exists j$, and $c_1$ is a constant of integration. The latter corresponds to the Gauss-Bonnet term in the Lagrangian whilst the others are the non-trivial solutions. Trivially, the vacuum solution is also a solution since $A_i = B_j = 0 \; \forall i$ and $f(0,0) = 0$, although this leaves the Lagrangian to be the Gauss-Bonnet term only, which does not contribute to the Friedmann equation and hence cannot be a source to the bounce. Thus, a fluid must exist. In this case, the vacuum condition is satisfied provided that any fluid obeying the $A_i$ condition satisfies
\begin{equation}
1+\dfrac{3\alpha}{2}(1+w_i) > 0 \implies w_i > \dfrac{-2-3\alpha}{3\alpha} \; \forall i.
\end{equation}

\subsection{$f(T,T_G) = -T + T_G g(T)$}

In this case, we enforce the presence of the TEGR term. This yields the following Friedmann equation
\begin{equation}
T- \dfrac{4T^3}{3}g_T = T_0 \sum\limits_i \Omega_{w_i,0} \left(\dfrac{T_0}{T}\right)^{-3(1+w_i)\alpha/2},
\end{equation} 
which yields the same solutions found in the previous section with an extra particular solution of the form
\begin{equation}
g_\text{part.}(T) = -\dfrac{3}{4T}.
\end{equation}
This introduces an extra contribution in the Lagrangian of the form $f_\text{part.}(T,T_G) = -\dfrac{3T_G}{4T}$. The vacuum conditions are identical to those found in the previous model since the new contributions reduce to zero in the $T \rightarrow 0$ limit.

\subsection{$f(T,T_G) = -T + \mu\left(\dfrac{T}{T_0}\right)^\beta\left(\dfrac{T_G}{T_{G,0}}\right)^\gamma$}

For this model, the Friedmann equation becomes
\begin{align}
&T + \mu\left(\dfrac{T}{T_0}\right)^\beta\left(\dfrac{T_G}{T_{G,0}}\right)^\gamma\left\lbrace 1 - 2\beta - \gamma-\dfrac{2\beta\gamma}{\alpha-1}- \dfrac{4\gamma(\gamma-1)}{\alpha-1} \right\rbrace \nonumber\\
&= T_0 \sum\limits_i \Omega_{w_i,0} \left(\dfrac{T_0}{T}\right)^{-3(1+w_i)\alpha/2}.
\end{align} 
The constant $\mu$ can be found by evaluating the expression at $t_* = t_0$ resulting in
\begin{align}
\mu= \dfrac{T_0 \left(-1+\sum\limits_i \Omega_{w_i,0}\right)}{1 - 2\beta - \gamma-\dfrac{2\beta\gamma}{\alpha-1}- \dfrac{4\gamma(\gamma-1)}{\alpha-1}},
\end{align} 
provided that the denominator is non-zero. This simplifies the Friedmann equation to
\begin{align}\label{eq:bouncingIV-power-friedmann}
&\dfrac{T}{T_0} + \left(\dfrac{T}{T_0}\right)^\beta\left(\dfrac{T_G}{T_{G,0}}\right)^\gamma  \left(-1+\sum\limits_i \Omega_{w_i,0}\right) = \sum\limits_i \Omega_{w_i,0} \left(\dfrac{T_0}{T}\right)^{-3(1+w_i)\alpha/2}.
\end{align} 
At this point, we consider two distinct scenarios, $\alpha = 1$ and $\alpha \neq 1$. In the former case, $T_G = 0$ at all times. Thus, the ratio of $T_G/T_{G,0}$ is not properly defined in this instance. Nonetheless, since $T_0$ and $T_{G,0}$ are constants, one can alternatively define a Lagrangian of the form $f(T,T_G) = -T + \nu T^\beta {T_G}^\gamma$, for some constant $\nu$. The Lagrangian is defined provided $\gamma > 0$ (and by the vacuum condition, provided that $\beta \geq 0$). In this case, the field equation reduces to
\begin{equation}
T + \nu T^\beta {T_G}^\gamma \left[1 - 2 \beta - \gamma - \dfrac{4T^2}{3T_G}\gamma\left(\beta+2\gamma-2\right)\right]  = T_0 \sum\limits_i \Omega_{w_i,0} \left(\dfrac{T_0}{T}\right)^{-3(1+w_i)/2}.
\end{equation} 
For the field equation to give physical results, one needs to further restrict the parameters $\beta$ and $\gamma$. The following cases are generated. If either $\gamma > 1$  or $\beta = 2-2\gamma$ (and since $\beta \geq 0$ and $\gamma > 0$, this restricts $0 < \gamma \leq 1$), the equation simplifies to
\begin{equation}
1 = \sum\limits_i \Omega_{w_i,0} \left(\dfrac{T_0}{T}\right)^{\frac{-3(1+w_i)+2}{2}}.
\end{equation} 
Since the LHS is a constant, the Friedmann equation is satisfied only when there exists a single fluid with EoS $w = -1/3$. Lastly, if $\gamma = 1$, the Friedmann equation simplifies to
\begin{equation}\label{eq:BouncingIV-power-GBfriedmann}
T - \dfrac{4}{3}\beta \nu T^{\beta+2}  = T_0 \sum\limits_i \Omega_{w_i,0} \left(\dfrac{T_0}{T}\right)^{-3(1+w_i)/2}.
\end{equation} 
By evaluating the expression at $t_* = t_0$, the value of $\nu$ can be found, being
\begin{equation}
\nu = \dfrac{1-\sum\limits_i \Omega_{w_i,0}}{\dfrac{4}{3}\beta {T_0}^{\beta+1}},
\end{equation} 
which is defined when $\beta > 0$. Assuming this is the case, the Friedmann equation can be expressed as
\begin{equation}
1 = \left(1-\sum\limits_i \Omega_{w_i,0}\right)\left(\dfrac{T}{T_0}\right)^{\beta+1} + \sum\limits_i \Omega_{w_i,0} \left(\dfrac{T_0}{T}\right)^{\frac{-3(1+w_i)+2}{2}}.
\end{equation} 
Since the LHS is a constant, the time (torsional) dependent components must cancel. Irrespective whether in vacuum or fluids exist, the condition $\beta = -1$ must be satisfied which originates from the first term on the LHS. However, this does not obey the vacuum condition $f(0,0) = 0$ since it requires $\beta \geq 0$. Now, if we consider $\beta = 0$, this would correspond to a Gauss-Bonnet contribution. However, from Eq. \ref{eq:BouncingIV-power-GBfriedmann}, this is only possible provided that a fluid exists with EoS $w = -1/3$. In fact, the result agrees with the case when $\beta = 2-2\gamma$ since when $\gamma = 1$, $\beta = 0$. \newline

For the case when $\alpha \neq 1$, the Friedmann equation Eq. \ref{eq:bouncingIV-power-friedmann} can be expressed in terms of time as
\begin{align}
&\left(\dfrac{t_0}{t_*}\right)^2 + \left(\dfrac{t_0}{t_*}\right)^{2\beta+4\gamma}  \left(-1+\sum\limits_i \Omega_{w_i,0}\right) = \sum\limits_i \Omega_{w_i,0} \left(\dfrac{t_0}{t_*}\right)^{3(1+w_i)\alpha}.
\end{align} 
The expression is satisfied for all times when the powers of $t_*$ cancel, leading to the following conditions
\begin{align}
\beta + 2\gamma &= 1 \\
3(1+w_i)\alpha &= 2, \, \forall i.
\end{align}
The first condition restricts the powers of $\beta$ and $\gamma$ whilst the second restricts the possible choice of fluids depending on the value of $\alpha$. In the case of vacuum, the second condition is not present. One can easily conclude that, in a non-vacuum universe, since all fluids must satisfy the second condition, the only possibility is that only one fluid is present (i.e. two fluids with different EoS parameters is not achievable). This reduces the problem to a standard single fluid dominated universe (unless vacuum is considered). Furthermore, since $\alpha > 0$, the range of EoS parameter values is restricted within $w > -1$. \newline

Lastly, given that the denominator of $\mu$ has to be non-zero, we get an extra condition being that 
\begin{equation}
\gamma \neq \dfrac{\alpha-1}{3\alpha-1},
\end{equation} 
whilst the vacuum solution condition demands $\beta + 2\gamma > 0$, which is ensured by the first condition.

\section{Bouncing model IV: Critical density}
\label{model4}

For this bouncing model, the scale factor takes the form
\begin{equation}
a(t) = A\left(\dfrac{3}{2}\rho_\text{cr}t^2+1\right)^{1/3},
\end{equation}
where $\rho_\text{cr}$ is the critical density and $A > 0$ is a dimensionless constant, which is the value of the scale factor at $t = 0$ i.e. $A = a(0)$. In this case, we find
\begin{align}
H &= \dfrac{2t \rho_\text{cr}}{2+3t^2 \rho_\text{cr}}, & T &= 6H^2 = 6\left(\dfrac{2t \rho_\text{cr}}{2+3t^2 \rho_\text{cr}}\right)^2, & T_G &= \dfrac{T^2}{3}\left(\dfrac{2}{t^2 \rho_\text{cr}}-1\right).
\end{align}
Here, the bounce occurs at $t = 0$ since $H(t<0) < 0$, $H(t=0) = 0$ and $H(t > 0) > 0$. Let us first express the scale factor and $T_G$ solely in terms of $T$. This can be achieved by expressing the time parameter $t$ in terms of $H$. From the definition of $H$, we have
\begin{equation}
3t^2 \rho_\text{cr}H- 2t \rho_\text{cr}+2H = 0,
\end{equation}
which is a quadratic in $t$ whose solution is
\begin{equation}
3Ht = 1 - \sqrt{1-\dfrac{6H^2}{\rho_\text{cr}}}.
\end{equation}
The correct sign was obtained by evaluating the expression at $t = 0$ since for $t = 0$, $H = 0$ thus leaving the negative sign as the physical solution. Therefore, the scale factor can be expressed in terms of $T$ as
\begin{equation}
a(T) = A\left[\dfrac{2\rho_\text{cr}}{T}\left(1 - \sqrt{1-\dfrac{T}{\rho_\text{cr}}}\right)\right]^{1/3},
\end{equation}
whilst the TEGB term is given by
\begin{equation}
T_G = -\dfrac{4T^2}{3} + 2T \rho_\text{cr}  \left(1+\sqrt{1-\frac{T}{\rho_\text{cr}}}\right).
\end{equation}
We also remark that the square root is always real. From the definition of $H$, one can easily find that the maximum value is achieved at the maximum turning point(s) which occurs at $t_{\text{max.}} = \pm\sqrt{\frac{2}{3\rho_{cr}}}$ being $H_{\text{max.}} = \pm\sqrt{\frac{\rho_{cr}}{6}}$. Thus, the maximum value for the torsion scalar is $T_\text{max.} = \rho$. Consequently, this leads to $0 \leq T/\rho_{cr} \leq 1$. In addition, in order to simplify the field equations and express them to be compared to observational data, we define an the current time $t_0 > 0$ where $a(t_0) = 1$,
\begin{equation}
{t_0}^2 = \dfrac{2}{3\rho_\text{cr}}\left(\dfrac{1}{A^3}-1\right).
\end{equation}
Since $\rho_\text{cr} > 0$, this equation holds provided that $A < 1$, which will be assumed from here thereon. Then, the parameters $T_0 \equiv T(t = t_0) = 4A^3(1-A^3)\rho_\text{cr}$ and $\Omega_{w_i,0} \equiv \Omega_{w_i}(t = t_0)$ provide their values at the current time.

\subsection{$f(T,T_G) = g(T) + h(T_G)$}

For this type of model, the Friedmann equation becomes
\begin{align}
&g + h - 2T g_T - T_G h_{T_G} + \frac{2T}{9} \left(-20 T^3+12 \rho  T^2-51 T T_G+36 \rho T_G \right) h_{T_{G}T_G} \nonumber \\
&= T_0 \sum\limits_i \Omega_{w_i,0} A^{-3(1+w_i)}\left[\dfrac{2\rho_\text{cr}}{T}\left(1 - \sqrt{1-\dfrac{T}{\rho_\text{cr}}}\right)\right]^{-(1+w_i)}. \label{modelV1}
\end{align} 
By using the above expressions for $T$ and $T_G$, the following relation is found:
\begin{equation}
T = \frac{3 {\rho_\text{cr}}}{16}-\frac{1}{32} \sqrt{96 T_G \left(\frac{9{\rho_\text{cr}}}{\sqrt{x}}-8\right)+27 \rho ^2 \left(\frac{{\rho_\text{cr}}}{\sqrt{x}}+4\right)-256 x}+\frac{\sqrt{x}}{2},
\label{modelV4}
\end{equation}
where 
\begin{align}
x &\equiv \frac{9 {\rho_\text{cr}}^2}{64}+\frac{1}{8} \sqrt[3]{-512 {T_G}^3+1161 {\rho_\text{cr}}^2 {T_G}^2+9 \sqrt{3} \sqrt{-3072 {\rho_\text{cr}}^2 {T_G}^5+4523 {\rho_\text{cr}}^4 {T_G}^4+192 {\rho_\text{cr}}^6 {T_G}^3}}\nonumber \\
&+\frac{16 {T_G}^2-9 {\rho_\text{cr}}^2 {T_G}}{2 \sqrt[3]{-512 {T_G}^3+1161 {\rho_\text{cr}}^2 {T_G}^2+9 \sqrt{3} \sqrt{-3072 {\rho_\text{cr}}^2 {T_G}^5+4523 {\rho_\text{cr}}^4 {T_G}^4+192 {\rho_\text{cr}}^6 {T_G}^3}}}-{T_G}.
\end{align}
Thus, equation (\ref{modelV1}) can be split in the following system of equations:
\begin{align}
&g - 2T g_T = T_0 \sum\limits_i \Omega_{w_i,0} A^{-3(1+w_i)}\left[\dfrac{2\rho_\text{cr}}{T}\left(1 - \sqrt{1-\dfrac{T}{\rho_\text{cr}}}\right)\right]^{-(1+w_i)}, \label{modelV2} \\
&h - T_G h_{T_G} + \frac{2T}{9} \left(-20 T^3+12 {\rho_\text{cr}}T^2-51 T T_G+36 {\rho_\text{cr}} T_G \right) h_{T_{G}T_G} = 0. \label{modelV3}
\end{align}
whose solution for $g(T)$ yields:
\begin{align}
g(T) &= c_1 \sqrt{T} -\frac{\Omega_{0,0} \sqrt{T} T_0}{4 A^3}\left[\frac{2 \sqrt{T}}{{\rho_\text{cr}} \left(\sqrt{1-\frac{T}{{\rho_\text{cr}}}}-1\right)}+\frac{2 \tan ^{-1}\left(\sqrt{\dfrac{{\rho_\text{cr}}}{T}-1}\right)}{\sqrt{{\rho_\text{cr}}}}\right] \nonumber \\
&+ \sum\limits_i \frac{\Omega_{w_i,0} T_0}{2 w_i} A^{-3 (1+w_i)} \left(\sqrt{1-\frac{T}{{\rho_\text{cr}}}}+1\right) \Bigg[(1+w_i) \, _2F_1\left(-\frac{1}{2},w_i;\frac{1}{2};1-\frac{2}{\sqrt{1-\frac{T}{{\rho_\text{cr}}}}+1}\right)\nonumber \\
&-\left(\frac{2}{\sqrt{1-\frac{T}{{\rho_\text{cr}}}}+1}\right)^{-w_i}\Bigg],
\end{align}
where $c_1$ is an integration constant corresponding to the DGP term. Note that in the case of dust ($w = 0$) has a distinct solution due to the divergence present in the summation. In this case, the vacuum condition implies
\begin{equation}
g(0) = \sum\limits_j \Omega_{w_j,0} T_0 A^{-3 (1+w_j)},
\end{equation}
where the summation includes the matter fluid.

The solution for $h(T_G)$ turns out more difficult to be obtained analytically, as the equation (\ref{modelV3}) together with the expression (\ref{modelV3}) requires numerical resources. Moreover, vacuum $f(0,0) = g(0) + h(0) = 0$ is only achieved in absence of matter fluids for $g(T)$ while the absence of an analytical solution for $h(T_G)$ prevents to go further with this analysis.

\subsection{$f(T,T_G) = T g(T_G)$}

For a $T$ rescaling model for some function $g(T_G)$, the Friedmann equation simplifies to
\begin{align}
&g +\left(\dfrac{4T^2}{3} - T_G\right) g_{T_G} -  \frac{2T}{9} g_{T_{G}T_G} \left(-20 T^3+12 \rho_\text{cr}  T^2-51 T T_G+36 \rho_\text{cr}  T_G\right) \nonumber \\
&= -\dfrac{T_0}{T} \sum\limits_i \Omega_{w_i,0} A^{-3(1+w_i)}\left[\dfrac{2\rho_\text{cr}}{T}\left(1 - \sqrt{1-\dfrac{T}{\rho_\text{cr}}}\right)\right]^{-(1+w_i)}.
\end{align} 
Let us rewrite this equation by defining the variable $x \equiv \sqrt{1-\dfrac{T}{\rho_\text{cr}}}$, what yields 
\begin{align}
& -3 \left(x^2-1\right)^2 \left[x (2 x+1) (7 x-2)-3\right]^2 g''(x) \nonumber \\
&+\left(x^2-1\right) \left(332 x^7+778 x^6-1036 x^5-1013 x^4+164 x^3+388 x^2-36 x-9\right) g'(x) \nonumber \\
&+\left(8 x^2+x-3\right) \left[x (2 x+1) (7 x-2)-3\right]^2 g(x) \nonumber \\
&= \sum\limits_i\frac{\xi_i}{\left(x^2-1\right)} \left(8 x^2+x-3\right) \left[x (2 x+1) (7 x-2)-3\right]^2 \left(1+x\right)^{1+w_i}.
\label{666}
\end{align}
where $\xi_i \equiv \frac{\Omega_{w_i,0} T_0}{\rho_\text{cr}} 2^{-1-w_i} A^{-3 (1+w_i)}$.

The solution for the homogeneous part of equation (\ref{666}) is given by
\begin{align}
g(x) &= c_1 \left(1 - \dfrac{1}{2}x^2 + \frac{15}{648}x^4+\frac{1851}{2430}x^5 + \dots\right) \nonumber \\
&+ c_2\left(x + \frac{a_1}{6}x^2 -\frac{769 a_1}{648}x^4 + \frac{1706}{2430}x^5 + \dots \right).
\end{align}
For the vacuum condition, we require $f(0,0) = 0$. In this case, after multiplying the homogeneous solution by the torsion scalar, the condition is satisfied. Nevertheless, the general solution can not be found analytically, since the the RHS of the equation (\ref{666}) is not necessarily a polynomial, depending on $w_i$. Furthermore, using the Wronskian and Green's function method is not feasible either since neither homogeneous solution is expressed analytically in terms of some known function. Nonetheless, the homogeneous solutions correspond to the vacuum solution which satisfy the vacuum condition.

\subsection{$f(T,T_G) = T_G g(T)$}

For a $T_G$ rescaling model, the Friedmann equation is given by
\begin{equation}
- \dfrac{4T^3}{3}g_T  = T_0 \sum\limits_i \Omega_{w_i,0} A^{-3(1+w_i)}\left[\dfrac{2\rho_\text{cr}}{T}\left(1 - \sqrt{1-\dfrac{T}{\rho_\text{cr}}}\right)\right]^{-(1+w_i)}.
\end{equation} 
By defining the variable $x \equiv 1+\sqrt{1-\dfrac{T}{\rho_\text{cr}}}$, the equation becomes:
\begin{equation}
(2-x)^3 g_x = (1-x) \sum\limits_i \xi_{w_i} x^{w_i-2},
\end{equation} 
where $\xi_{w_i} \equiv -\frac{3T_0}{2{\rho_\text{cr}}^2} \Omega_{w_i,0} 2^{-(1+w_i)} A^{-3(1+w_i)}$. The general solution is given by:
\begin{align}
g(x) &= c_1+\frac{1}{16} \xi_w  x^w \bigg[\frac{2}{(w-1) x}+\frac{1}{w}+\frac{4 \left(\frac{x}{x-2}\right)^{-w} \, _2F_1\left(2-w,-w;3-w;-\frac{2}{x-2}\right)}{(w-2) (x-2)^2}\nonumber \\
&-\frac{\left(\frac{x}{x-2}\right)^{-w} \, _2F_1\left(-w,-w;1-w;-\frac{2}{x-2}\right)}{w}\bigg]. \label{modelVS1}
\end{align}
which diverges for dust $w=0$. For the case of a pressureless fluid, the solution reduces to: 
\begin{equation}
g(x) = c_1 + \frac{\xi_0}{16} \left[-\frac{2}{(x-2)^2}-\frac{2}{x}+\ln \left(\dfrac{x}{2-x}\right)\right],
\end{equation}
Nevertheless, such Lagrangians diverges in vacuum, where $T=T_G=0$. However, by assuming more than a single fluid, the general solution leads to the sum of the solutions (\ref{modelVS1}) for each EoS $w$, and vacuum may be achieved by the cancelation of the divergences. Particularly, by assuming an arbitrary number of fluids, the following condition is found
\begin{equation}
0 = \sum\limits_i a_i \xi_{w_i},
\end{equation}
where $a_i > 0$ are unknown coefficients corresponding to each EoS. However, as $a_i \xi_{w_i} < 0 \; \forall i$, and the solution does not describe the bouncing cosmology whilst obeying the vacuum condition.

\subsection{$f(T,T_G) = -T + T_G g(T)$}

For a $T_G$ rescaling with a TEGR contribution, the Friedmann equation becomes
\begin{equation}
T - \dfrac{4T^3}{3} g_T = T_0 \sum\limits_i \Omega_{w_i,0} A^{-3(1+w_i)}\left[\dfrac{2\rho_\text{cr}}{T}\left(1 - \sqrt{1-\dfrac{T}{\rho_\text{cr}}}\right)\right]^{-(1+w_i)}.
\end{equation} 
In this case, the solution is similar to the previous model with an extra particular solution of the form $g_{\text{part.}} = -3/4T$. Thus, the Lagrangian is given by
\begin{equation}
f(T,T_G) = -T - \dfrac{3T_G}{4T} + T_G h(T),
\end{equation}
where $h(T)$ represents the previous model solution. To satisfy the vacuum condition, we again require $f(0,0) = 0$. However, as indicated in the previous model, $T_G h(T)|_{T, T_G \rightarrow 0}$ yields finite results only in vacuum. This leads to $f(0,0) = -3 \rho_\text{cr} < 0$. Therefore, this model does not satisfy the vacuum condition.

\subsection{$f(T,T_G) = -T + \mu\left(\dfrac{T}{T_0}\right)^\beta\left(\dfrac{T_G}{T_{G,0}}\right)^\gamma$}
For a power-law model, the Friedmann equation becomes
\begin{align}
&T_0 \sum\limits_i \Omega_{w_i,0} a^{-3(1+w_i)} = T + \mu\left(\dfrac{T}{T_0}\right)^\beta\left(\dfrac{T_G}{T_{G,0}}\right)^\gamma  \bigg[1- 2\beta - \gamma + \beta \gamma \left(2-\frac{4 T^2}{3 T_G}\right) \nonumber \\
&+\gamma(\gamma-1)\frac{2 T}{9 {T_G}^2} \left(-20 T^3+12 \rho_\text{cr}T^2-51 T {T_G}+36 \rho_\text{cr} {T_G}\right)\bigg],
\end{align}  
For this model, the vacuum condition $f(0,0) = 0$ is satisfied as long as $\beta + \gamma > 0$. Evaluating the Friedmann equation at $t = 0$ yields the following condition,
\begin{equation}
0 = \sum\limits_i \Omega_{w_i,0} A^{-3(1+w_i)}.
\end{equation}
However, since both parameters are positive, this is not achievable unless vacuum is considered. Thus, the latter is assumed. By evaluating the Friedmann equation at $t = t_0$, the constant $\mu$ can be determined to be
\begin{align}
\mu &= \dfrac{-T_0}{1- 2\beta - \gamma + \beta \gamma \left(2-\frac{4 {T_0}^2}{3 T_{G,0}}\right)+\frac{2 \gamma(\gamma-1) T_0}{9 {T_{G,0}}^2} \left(-20 {T_0}^3+12 \rho_\text{cr}{T_0}^2-51 T_0 {T_{G,0}}+36 \rho_\text{cr} {T_{G,0}}\right)} \nonumber \\
&\equiv -\dfrac{T_0}{\nu},
\end{align}
where $\nu \neq 0$ is defined to be the denominator. This simplifies the Friedmann equation to be
\begin{align}
&T = \dfrac{T_0}{\nu}\left(\dfrac{T}{T_0}\right)^\beta\left(\dfrac{T_G}{T_{G,0}}\right)^\gamma  \bigg[1- 2\beta - \gamma + \beta \gamma \left(2-\frac{4 T^2}{3 T_G}\right) \nonumber \\
&+\gamma(\gamma-1)\frac{2 T}{9 {T_G}^2} \left(-20 T^3+12 \rho_\text{cr}T^2-51 T {T_G}+36 \rho_\text{cr} {T_G}\right)\bigg],
\end{align} 
Since we require the equation to hold at all times, assuming $T \neq 0$, the Friedmann equation can be rearranged to be in the form $\nu = g(T)$,
for some function $g$. Thus, since the LHS is a constant, the RHS must also be a constant meaning that the function must be independent of $T$. This is true under two cases, $\beta = -1, \gamma = 1$ and $\beta = 1, \gamma = 0$. The former, albeit leading to a non-trivial Lagrangian, does not satisfy the vacuum condition. On the other hand, the second case corresponds to a TEGR rescaling with $\nu = -1$. However, this leads to a zero Lagrangian which is non-physical. Therefore, this case is also neglected.\newline

We conclude this section by examining the TEGR with DGP and Gauss-Bonnet terms since the latter two do not contribute to the Friedmann equation. In this case, the equation becomes
\begin{align}
&T = T_0 \sum\limits_i \Omega_{w_i,0} a^{-3(1+w_i)}.
\end{align}  
At time $t = 0$, the same condition is obtained, which is only true when vacuum is considered. However, this would imply that $T = 0$ at all times which is a contradiction. Thus, this implies that the TEGR term cannot describe the bouncing cosmology. Therefore, no Lagrangian has been found which satisfies the vacuum condition.



\section{Bouncing Model V: Exponential Model II}
\label{model5}

The last bouncing model we analysed here, it is similar to the first one, but it may include a future singularity, similar to the power-law model studied above:
\begin{equation}
a(t) = A \exp \left[\dfrac{f_0}{\alpha+1}(t-t_s)^{\alpha+1}\right],
\end{equation}
where $A > 0$ is a dimensionless constant which corresponds to the scale factor at the bouncing point time $t_s$ i.e. $A = a(t_s)$, $f_0 > 0$ is some arbitrary  constant having time dimensions [T]$^{-\alpha-1}$ and $\alpha$ is a constant. In this case, the Hubble parameter, and consequently the torsion scalar and TEGB term are given by
\begin{align}
H &= f_0 (t-t_s)^{\alpha}, & T &= 6H^2, & T_G &= 4T\left[\dfrac{T}{6} + f_0 \alpha \left(\dfrac{T}{6{f_0}^2}\right)^{\frac{\alpha-1}{2\alpha}}\right].
\end{align}
Furthermore, the scale factor can be solely expressed in terms of the torsion scalar as
\begin{equation}
a(T) = A \exp \left[\dfrac{f_0}{\alpha+1}\left(\dfrac{T}{6{f_0}^2}\right)^{\frac{\alpha+1}{2\alpha}}\right],
\end{equation}
A type IV singularity (\cite{Nojiri:2005sx}) may occur in this bouncing cosmology when
\begin{equation}\label{eq:alphacond_bouncingVI}
\alpha = \dfrac{2n+1}{2m+1},
\end{equation}
where $n, m \in \mathbb{N}$ and $\alpha > 1$. Before reconstructing the corresponding Lagrangians, we make note that by introducing the new time variable $t_* \equiv t - t_s$, the scale factor and Hubble parameter become
\begin{align}
a(t_*) &= A \exp \left[\dfrac{f_0}{\alpha+1}{t_*}^{\alpha+1}\right], & H &= f_0 {t_*}^{\alpha}.
\end{align}
This effectively simplifies the Hubble parameter to be a standard power law relation in the time variable $t_*$. 
Lastly, we define an instant of time $t_* = t_0 > 0$ at which $a(t_0) = 1$ to simplify the Friedmann equation's calculations. The time is given by
\begin{equation}\label{eq:current-time_modelVI_rel}
{t_0}^{\alpha+1} = -\dfrac{\alpha+1}{f_0}\ln A.
\end{equation}
Since we demand that $t_0 > 0$, and $\alpha, f_0 > 0$, we require $0 < A < 1$. In what follows, this will be assumed. By defining this time, we define the torsion scalar at this instant as follows
\begin{equation}
T_0 \equiv T(t_* = t_0) = 6{f_0}^2 {t_0}^{2\alpha}.
\end{equation}
By doing so, the scale factor simplifies to
\begin{equation}
a(T) = A^{1-\left(\frac{T}{T_0}\right)^{\frac{\alpha+1}{2\alpha}}},
\end{equation}
where we have used Eq. (\ref{eq:current-time_modelVI_rel}). 
Furthermore, the TEGB term can be re-expressed into a simpler form as follows
\begin{equation}\label{eq:TGform-bouncingVI}
T_G = 4T\left[\dfrac{T}{6} +\alpha f_0 {t_0}^{\alpha-1} \left(\dfrac{T}{T_0}\right)^{\frac{\alpha-1}{2\alpha}}\right].
\end{equation}
However, working with this scale factor may introduce difficulties when reconstructing the corresponding gravitational actions. Instead, we make use of Eq. (\ref{eq:current-time_modelVI_rel}), such that the scale factor can be expressed as
\begin{equation}
a(T) = \exp \left\lbrace -\dfrac{f_0 {t_0}^{\alpha+1}}{\alpha+1} \left[1-\left(\frac{T}{T_0}\right)^{\frac{\alpha+1}{2\alpha}}\right]\right\rbrace.
\end{equation} 

\subsection{$f(T,T_G) = g(T) + h(T_G)$}

For a separable additional model for $T$ and $T_G$, the Friedmann equation reduces to
\begin{align}
&g + h - 2T g_T - T_G h_{T_G} -\frac{\left(2 T^2-3T_G\right) \left[2 (\alpha +1) T^2+3(3\alpha-1)T_G\right]}{9 \alpha} h_{T_{G}T_G} \nonumber \\
&= T_0 \sum\limits_i \Omega_{w_i,0} \exp \left\lbrace \dfrac{3f_0 {t_0}^{\alpha+1}(1+w_i)}{\alpha+1} \left[1-\left(\frac{T}{T_0}\right)^{\frac{\alpha+1}{2\alpha}}\right]\right\rbrace.
\end{align} 
This equation can not be split as previous cases due to the coefficient of $h_{T_G T_G}$. There may exist an invertible relation for $T$ in terms of $T_G$, such that $T = p(T_G)$, but not a general one for any arbitrary $\alpha$. Indeed, given the form of $\alpha$ in Eq. (\ref{eq:alphacond_bouncingVI}) with $\alpha > 1$, the form of $T_G$ is given as
\begin{equation}\label{eq:TG_secondform_bouncingVI}
T_G = \mu T^2 + \nu T^{\frac{3 n-m+1}{2 n+1}},
\end{equation}
where $\mu$ and $\nu$ are the corresponding coefficients of Eq. \ref{eq:TGform-bouncingVI}. It is clear that due to the last term, the equation is, in general, not invertible. Nonetheless, in some particular cases, the equation is invertible. For sake of generality, we assume that $T$ is invertible and some function $p(T_G)$ exists. In other words, the Friedmann equation now becomes
\begin{align}
&g + h - 2T g_T - T_G h_{T_G} -q(T_G) h_{T_{G}T_G} \nonumber \\
&= T_0 \sum\limits_i \Omega_{w_i,0} \exp \left\lbrace \dfrac{3f_0 {t_0}^{\alpha+1}(1+w_i)}{\alpha+1} \left[1-\left(\frac{T}{T_0}\right)^{\frac{\alpha+1}{2\alpha}}\right]\right\rbrace,
\end{align} 
where $q(T_G)$ is a function of the TEGB term only representing the coefficient of $h_{T_G T_G}$, which is now possible due to the demand that $T = p(T_G)$. Now, the equation can be separated with each side of the equation in terms of $T$ and $T_G$ independently, leading to the same procedure used in Section \ref{sec:addition-bouncingIV}. In fact, the constant which is generated can be set to zero as it will not contribute to the Lagrangian. Thus, the system of differential equations lead to
\begin{align}
&g - 2T g_T = T_0 \sum\limits_i \Omega_{w_i,0} \exp \left\lbrace \dfrac{3f_0 {t_0}^{\alpha+1}(1+w_i)}{\alpha+1} \left[1-\left(\frac{T}{T_0}\right)^{\frac{\alpha+1}{2\alpha}}\right]\right\rbrace, \\
&h - T_G h_{T_G} -q(T_G) h_{T_{G}T_G} = 0.
\end{align} 
The solution for $g(T)$ is given by
\begin{align}
&g(T) = c_1 \sqrt{T} \nonumber \\
&+ \sum_i\sum_{n=0}^{\infty } \frac{\Omega_{w_i,0} T_0}{n!} \left(\frac{3 f_0 {t_0}^{\alpha +1}(1+w_i)}{\alpha +1}\right)^n \, _2F_1\left[-n,-\frac{\alpha }{\alpha +1};\frac{1}{\alpha +1};\left(\frac{T}{T_0}\right)^{\frac{\alpha +1}{2 \alpha }}\right],
\end{align}
where $c_1$ is an integration constant whose term corresponds to the DGP term and $_2F_1(a,b,c;z)$ is Gauss' hypergeometric function. Note that since $\alpha > 1$, the hypergeometric function is always defined. When $T = 0$, the solution reduces to
\begin{equation}
g(0) = \sum\limits_i \Omega_{w_i,0} T_0 \exp \left[\frac{3 f_0 (1+w_i) {t_0}^{\alpha +1}}{\alpha +1}\right].
\end{equation}
As discussed at the beginning of this section, the form of $q(T_G)$ is unknown or non-existent depending on the value of $\alpha$. The exponent of the last term in Eq. (\ref{eq:TG_secondform_bouncingVI}) lies on the range $(1,3/2)$, leading to problems for getting an invertible condition. Nonetheless, equation generates two independent homogeneous solutions since it is a linear homogeneous type, say $u_1(T_G)$ and $u_2(T_G)$. Thus, the solution for $h$ can always be expressed as
\begin{equation}
h(T_G) = c_1 u_1(T_G) + c_2 u_2(T_G),
\end{equation}
for some arbitrary integration constants $c_{1,2}$. In fact, it is easy to verify that one of the solutions is the Gauss-Bonnet contribution $T_G$. In other words, the solution is 
\begin{equation}
h(T_G) = c_1 u_1(T_G) + c_2 T_G.
\end{equation}
Now, independently of the form of $u_{1}$, we can reach the following conclusions. If the function $u_{1}(0) = 0$, then this gives a non-trivial solution with $h(0) = 0$. This demands that $g(0) = 0$ for the vacuum condition to be satisfied, which is possible only in absence of matter. On the other hand, if this results into a constant, this still defines a non-trivial solution, however $h(0)$ can be non-zero depending on the integration constant. If the integration constant is set to zero, then $g(0) = 0$ which is only possible in vacuum. On the other hand, if $h(0)$ is equal to some constant $\mu \neq 0$, then $g(0) = -h(0) = -\mu$. Furthermore, since $g(0) > 0$ in these cases, this restricts $\mu < 0$. Lastly, if the function diverges at $T_G = 0$, the singularity can be removed by setting the integration constant to zero. Again, this sets $h(0) = 0$ leading $g(0) = 0$ for vacuum solutions to occur, which is again only satisfied in vacuum. 

\subsection{$f(T,T_G) = T g(T_G)$}

For a rescaling of $T$ model, the resulting Friedmann equation is given by
\begin{align}
&g + g_{T_G}\left(-T_G+\dfrac{4T^2}{3}\right) +\frac{\left(2 T^2-3T_G\right) \left[2 (\alpha +1) T^2+3(3\alpha-1)T_G\right]}{9 \alpha} g_{T_{G}T_G} \nonumber \\
&= -\dfrac{T_0}{T} \sum\limits_i \Omega_{w_i,0} \exp \left\lbrace \dfrac{3f_0 {t_0}^{\alpha+1}(1+w_i)}{\alpha+1} \left[1-\left(\frac{T}{T_0}\right)^{\frac{\alpha+1}{2\alpha}}\right]\right\rbrace.
\label{modelVIB1}
\end{align} 
Similar to the previous case, a problem arises due to the invertibility issue of the torsion scalar as a function of the TEGB term. Nonetheless, we can extract and analyse some behaviours of the solution even in absence of its explicit form. Let us express the equation (\ref{modelVIB1}) in terms of $T_G$:
\begin{equation}
g + p(T_G)g_{T_G} + q(T_G) g_{T_G T_G} = h(T_G),
\label{modelVIB2}
\end{equation}
where $p, q$ and $h$ are unknown functions pf $T_G$. Thus, the complete solution would be given by
\begin{equation}
g(T_G) =  c_1 u_1(T_G) + c_2 u_2(T_G) + \int^{T_G} G(T_G,s) h(s) {\rm d}s\ . \label{modelVIB3}
\end{equation}
where $G(T_G,s)$ is the Green function of the equation (\ref{modelVIB2}) while $u_{1,2}(T_G)$ are the solutions of the homogeneous part of the equation (\ref{modelVIB2}). Finally, the vacuum condition is satisfied, i.e. $T = T_G = 0$ implies $f(0,0) = 0$, as far as the solution (\ref{modelVIB3}) is finite at $T_G=0$.

\subsection{$f(T,T_G) = T_G g(T)$}

For a TEGB rescaling model, the resulting equation yields
\begin{equation}
- \dfrac{4T^3}{3}g_T  = T_0 \sum\limits_i \Omega_{w_i,0} \exp \left\lbrace \dfrac{3f_0 {t_0}^{\alpha+1}(1+w_i)}{\alpha+1} \left[1-\left(\frac{T}{T_0}\right)^{\frac{\alpha+1}{2\alpha}}\right]\right\rbrace.
\end{equation} 
The first solution of this equation is given by
\begin{equation}
g_1(T) = c_1 + \sum\limits_i \sum\limits_{n=0}^\infty \frac{3 \Omega_{w_i,0} T_0}{8 T^2 n!} \left[\dfrac{3f_0 {t_0}^{\alpha+1}(1+w_i)}{\alpha+1}\right]^n \, _2F_1\left[-n,-\frac{4 \alpha }{\alpha +1};1-\frac{4 \alpha }{\alpha +1};\left(\frac{T}{T_0}\right)^{\frac{\alpha +1}{2 \alpha }}\right],
\end{equation}
where $c_1$ is an integration constant, which corresponds to the Gauss-Bonnet contribution in the Lagrangian, and $_2F_1(a,b;c;z)$ is Gauss' hypergeometric function. The solution exists and is defined provided that the third argument in the hypergeometric function $c \equiv 1-\frac{4 \alpha }{\alpha +1} \not\in \mathbb{Z}^{-}\cup \lbrace 0 \rbrace$. For the values of $\alpha$ considered for the type IV singularity in Eq. (\ref{eq:alphacond_bouncingVI}) with $\alpha > 1$, the only allowed value for $\alpha = 3$ that results $c = -2$. This leads to the second solution
\begin{equation}
g_2(T) = c_1 + \sum\limits_i \frac{3 \Omega_{w_i,0} T_0}{16 T^2}\exp \left(\dfrac{3 f_0 {t_0}^4 (1+w_i)}{4}\right) \left[{\rm e}^{x_i}\left({x_i}^2+x_i+2\right)-{x_i}^3 \text{Ei}(x_i)\right] ,
\end{equation}
where $x_i \equiv -\dfrac{3 f_0 {t_0}^4 (1+w_i)}{4}\dfrac{T^{2/3}}{{T_0}^{2/3}}$ and $\text{Ei}(z)$ is the exponential integral. 
Whether both solutions satisfy $f(0,0) = 0$ can be checked by evaluating the solutions in vacuum:
\begin{align}
f(0,0) &= \sum\limits_i \frac{\Omega_{w_i,0} T_0}{4} \exp \left[ \dfrac{3f_0 {t_0}^{\alpha+1}(1+w_i)}{\alpha+1}\right] \nonumber \\
&+ \sum\limits_i \sum\limits_{n=0}^\infty\frac{3 \alpha f_0 \Omega_{w_i,0} T_0 {t_0}^{\alpha-1}}{2 T n!} \left(\frac{T}{{T_0}}\right)^{\frac{\alpha -1}{2 \alpha }} \left[\frac{3 f_0 {t_0}^{\alpha+1} (1+w_i)}{\alpha +1}\right]^n \nonumber \\
& \times \, _2F_1\left[-n,-\frac{4 \alpha }{\alpha +1};1-\frac{4 \alpha }{\alpha +1};\left(\frac{T}{{T_0}}\right)^{\frac{\alpha +1}{2 \alpha }}\right],
\end{align}
which gives a singularity in the second summation due to $\alpha > 1$ condition. Trivially, the condition is satisfied when vacuum is considered although this results a Lagrangian with only the Gauss-Bonnet term which is non-physical. On the other hand, the singularity can be removed only when all the coefficients sum to zero, i.e.
\begin{align}
0 &= \sum\limits_i \sum\limits_{n=0}^\infty\frac{3 \alpha f_0 \Omega_{w_i,0} T_0 {t_0}^{\alpha-1}}{2 n!} \left[\frac{3 f_0 (1+w_i) {t_0}^{\alpha+1}}{\alpha +1}\right]^n \nonumber \\
&= \sum\limits_i \frac{3 \alpha f_0 \Omega_{w_i,0} T_0 {t_0}^{\alpha-1}}{2} \exp \left[ \dfrac{3f_0 {t_0}^{\alpha+1}(1+w_i)}{\alpha+1}\right].
\end{align}
However, since every contribution is positive, the condition cannot be satisfied.

On the other hand, for the second solution, one finds
\begin{align}
f(0,0) &= \sum\limits_i \dfrac{\Omega_{w_i,0}}{2}\exp \left(\dfrac{3 f_0 {t_0}^4 (1+w_i)}{4}\right)\bigg\lbrace -\frac{T_0 (1+9 w_i)}{16} & \nonumber \\
&+9 f_0 {t_0}^2\left(\frac{T_0}{T}\right)^{2/3} \exp \left(-\frac{3}{4}f_0 {t_0}^4 (1+w_i) \frac{T^{2/3}}{{T_0}^{2/3}}\right)\bigg\rbrace\bigg|_{T \rightarrow 0},
\end{align}
which owns a singularity in the exponential term provided vacuum is not considered (in this case, the solution trivially holds although the Lagrangian would only be provided by the Gauss-Bonnet term which is non-physical). The singularity in the exponential term can be removed only if the coefficients sum to 0, i.e.
\begin{equation}
\sum\limits_i f_0 {t_0}^2 \Omega_{w_i,0} = 0.
\end{equation} 
However, since $f_0, t_0, \Omega_{w_i,0} > 0$, this condition cannot be satisfied leading to the vacuum solution as the only solution which satisfies the vacuum condition, as natural.

\subsection{$f(T,T_G) = -T + T_G g(T)$}

For models with a TEGB rescaling and a TEGR contribution, the resulting equation is
\begin{equation}
T - \dfrac{4T^3}{3} g_T = T_0 \sum\limits_i \Omega_{w_i,0} \exp \left\lbrace \dfrac{3f_0 {t_0}^{\alpha+1}(1+w_i)}{\alpha+1} \left[1-\left(\frac{T}{T_0}\right)^{\frac{\alpha+1}{2\alpha}}\right]\right\rbrace.
\end{equation} 
Here, the solutions are identical to the previous case with an extra particular solution
\begin{equation}
g_\text{part.}(T) = -\dfrac{3}{4T}.
\end{equation}
To check for vacuum solutions, we demand the condition $f(0,0) = 0$. Since the results in the previous section show that only vacuum can yield finite results in the $T, T_G \rightarrow 0$ limit, the resulting Lagrangian which must be checked for the vacuum condition is
\begin{equation}
f(T,T_G) = -T -\dfrac{3T_G}{4T} + c_1 T_G,
\end{equation}
where $c_1$ is a constant of integration. In this case, the limit does satisfy the vacuum condition and hence can describe the bouncing cosmology.

\subsection{$f(T,T_G) = -T + \mu\left(\dfrac{T}{T_0}\right)^\beta\left(\dfrac{T_G}{T_{G,0}}\right)^\gamma$}

For a power-law model in both $T$ and $T_G$, the Friedmann equation reduces to
\begin{align}
&T + \mu\left(\dfrac{T}{T_0}\right)^{\beta+\gamma\frac{3\alpha -1}{2 \alpha }}\left[\frac{6 \alpha  f_0 {t_0}^{\alpha -1}+T_0 \left(\frac{T}{T_0}\right)^{\frac{\alpha +1}{2 \alpha }}}{6 \alpha  f_0 {t_0}^{\alpha -1}+T_0}\right]^\gamma \Bigg\lbrace 1- 2\beta - \gamma + \frac{12 \beta\gamma \alpha  f_0 {t_0}^{\alpha-1}}{6 \alpha  {f_0} {t_0}^{\alpha-1}+ {T_0} \left(\frac{T}{T_0}\right)^{\frac{\alpha +1}{2 \alpha}}} \nonumber \\
&+\frac{12 \alpha \gamma(\gamma-1) f_0 {t_0}^{\alpha -1} \left[3 (3 \alpha -1) f_0 {t_0}^{\alpha -1}+2 T_0 \left(\frac{T}{T_0}\right)^{\frac{\alpha +1}{2 \alpha }}\right]}{\left[6 \alpha  f_0 {t_0}^{\alpha -1}+T_0 \left(\frac{T}{T_0}\right)^{\frac{\alpha+1}{2 \alpha }}\right]^2}\Bigg\rbrace = T_0 \sum\limits_i \Omega_{w_i,0} a^{-3(1+w_i)}.
\end{align}  
For this model, vacuum solutions are obtained provided that
\begin{equation}
\beta+\frac{(3 \alpha -1) \gamma }{2 \alpha } >0.
\end{equation}
The value of $\mu$ is obtained by evaluating the expression at current time, yielding
\begin{align}
&\mu= \dfrac{-T_0+T_0 \sum\limits_i \Omega_{w_i,0} a^{-3(1+w_i)}}{1- 2\beta - \gamma + \frac{12 \beta\gamma \alpha  f_0 {t_0}^{\alpha-1}}{6 \alpha  {f_0} {t_0}^{\alpha-1}+ {T_0}} +\frac{12 \alpha \gamma(\gamma-1) f_0 {t_0}^{\alpha -1} \left[3 (3 \alpha -1) f_0 {t_0}^{\alpha -1}+2 T_0 \right]}{\left(6 \alpha  f_0 {t_0}^{\alpha -1}+T_0 \right)^2}} \nonumber \\
&\equiv \frac{1}{\nu}\left(-T_0+T_0 \sum\limits_i \Omega_{w_i,0} a^{-3(1+w_i)}\right).
\end{align}  
where $\nu$ is defined by the denominator provided that it is non-zero. Note that the DGP ($\beta = 1/2, \gamma = 0$) and Gauss-Bonnet ($\beta = 0, \gamma = 1$) contributions cases give $\nu = 0$ and hence are excluded for the subsequent analysis. The special case when these are considered is discussed at the end of the section. Furthermore, by evaluating the expression at the bouncing time $t = t_s$ (or equivalently, $t_* = 0$), results in the following condition
\begin{equation}
\sum\limits_i \Omega_{w_i,0} A^{-3(1+w_i)} = 0.
\end{equation}
This condition can only be satisfied in absence of any type of matter, i.e. $\Omega_{w_i,0}=0$. Let us assume such a case, the Friedmann equation is simplified as follows 
\begin{align}
&\dfrac{T_0}{\nu}\left(\dfrac{T}{T_0}\right)^{\beta+\gamma\frac{3\alpha -1}{2 \alpha }}\left[\frac{6 \alpha  f_0 {t_0}^{\alpha -1}+T_0 \left(\frac{T}{T_0}\right)^{\frac{\alpha +1}{2 \alpha }}}{6 \alpha  f_0 {t_0}^{\alpha -1}+T_0}\right]^\gamma \Bigg\lbrace 1- 2\beta - \gamma + \frac{12 \beta\gamma \alpha  f_0 {t_0}^{\alpha-1}}{6 \alpha  {f_0} {t_0}^{\alpha-1}+ {T_0} \left(\frac{T}{T_0}\right)^{\frac{\alpha +1}{2 \alpha}}} \nonumber \\
&+\frac{12 \alpha \gamma(\gamma-1) f_0 {t_0}^{\alpha -1} \left[3 (3 \alpha -1) f_0 {t_0}^{\alpha -1}+2 T_0 \left(\frac{T}{T_0}\right)^{\frac{\alpha +1}{2 \alpha }}\right]}{\left[6 \alpha  f_0 {t_0}^{\alpha -1}+T_0 \left(\frac{T}{T_0}\right)^{\frac{\alpha+1}{2 \alpha }}\right]^2}\Bigg\rbrace = T.
\end{align}  
By assuming $T \neq 0$ (which already trivially satisfies the relation) gives
\begin{align}
&\left(\dfrac{T}{T_0}\right)^{\beta+\gamma\frac{3\alpha -1}{2 \alpha }-1}\left[\frac{6 \alpha  f_0 {t_0}^{\alpha -1}+T_0 \left(\frac{T}{T_0}\right)^{\frac{\alpha +1}{2 \alpha }}}{6 \alpha  f_0 {t_0}^{\alpha -1}+T_0}\right]^\gamma \Bigg\lbrace 1- 2\beta - \gamma + \frac{12 \beta\gamma \alpha  f_0 {t_0}^{\alpha-1}}{6 \alpha  {f_0} {t_0}^{\alpha-1}+ {T_0} \left(\frac{T}{T_0}\right)^{\frac{\alpha +1}{2 \alpha}}} \nonumber \\
&+\frac{12 \alpha \gamma(\gamma-1) f_0 {t_0}^{\alpha -1} \left[3 (3 \alpha -1) f_0 {t_0}^{\alpha -1}+2 T_0 \left(\frac{T}{T_0}\right)^{\frac{\alpha +1}{2 \alpha }}\right]}{\left[6 \alpha  f_0 {t_0}^{\alpha -1}+T_0 \left(\frac{T}{T_0}\right)^{\frac{\alpha+1}{2 \alpha }}\right]^2}\Bigg\rbrace = \nu.
\end{align}  
Since the LHS is constant, all the torsion terms on the RHS must vanish and yield a constant. This is possible only if $\beta = 1$ and $\gamma = 0$. This sets $\nu = -1$, so the Lagrangian turns out zero, which is not physical. \\



\section{Conclusions}
\label{Conclusions}
Bouncing cosmologies have become a reliable alternative to the inflationary paradigm, specially because the absence of initial conditions to start the cosmological evolution and also because the absence of an initial singularity within some models. In general, such scenario results in a universe that expands and then slows down and contracts again, a similar framework to the so-called ekpyrotic universes. Here we have investigated the possibility of reproducing some bouncing cosmologies in the framework of a class of extended Teleparallel theories, where the gravitational action includes functions of the torsion scalar and an analogous of the Gauss-Bonnet invariant. To do so, we have considered some particular forms of the Lagrangian according to some physical properties. \\

Then, several bouncing cosmologies have been considered, including some singular bouncing solutions, and the corresponding Lagrangian is reconstructed. Also the existence of vacuum (null torsion) solutions has been analysed, since it guarantees that such Lagrangians will indeed contain both Minkowski and Schwarzschild solutions, a fundamental requirement for the viability of any theory of gravity. Let us now summarise the solutions explored along the paper.  Firstly, we have considered a class of exponential law for the scale factor, free of singularities, where the scale factor decreases and  reaches a minimum, avoiding the occurrence of Big Bang-like singularity and then, increases. The Hubble parameter is then described by a linear function of the cosmic time, as shown in the first row of Fig.~\ref{Model1}. Despite this is not a realistic example, it represents quite well the idea of a bounce in the universe expansion. By considering several forms of the gravitational action, the corresponding function of the torsion scalar and the Gauss-Bonnet invariant is reconstructed. As shown in Section \ref{model1}, the analytical expression for the gravitational Lagrangian is difficult to be obtained but in general the action fulfills the requirement of vacuum solutions. Also an oscillating bouncing universe is considered. Such example is not regular for the whole cosmological history but contains a singularity, a Big Bang/Crunch singularity, such that the scale factor goes to zero and then the universe stars in a Big Bang again. Nevertheless, note that such singularity may be alleviated by imposing a minimum value larger than zero on the scale factor. The reconstructed Lagrangians corresponding to this oscillating solution are provided in Section \ref{model2}, although in general, the Lagrangians do not behave well in vacuum, where some of the reconstructed functions diverge. Then, a similar solution in terms of the occurrence of a Big Bang/Crunch singularity is also given in the form of a  power-law solution in Sect.~\ref{model3}. This case makes the gravitational action simpler for some of the classes of Lagrangians explored in the paper. In addition, vacuum solutions are better achieved for the power-law solution than in the previous case. Another important bouncing solution widely explored in the literature is the so-called Critical density solution, which is free of singularities and very similar to the exponential case in spite of exhibits a more complex - and realistic - evolution of the Hubble parameter. Nevertheless, the reconstruction of the corresponding Lagrangians turns out more difficult than in the previous cases, and only some analytical expressions are obtained, as shown in Sect.~\ref{model4}. Finally, we have explored an extension of the first model, the exponential case, with the presence of a possible future singularity. The corresponding discussion about the gravitational Lagrangians is raised in Section \ref{model5}, but in general the action becomes very complex and the analysis of vacuum solutions turns out not possible.    \\

Hence, we have explored a wide range of bouncing solutions in the framework of $f(T,T_G)$ actions, such that the corresponding Lagrangians can be reconstructed. Here, we have thus provided some techniques and tools for the analysis of this type of Lagrangians when analysing such cosmological solutions. Thus, we have shown the viability of some Lagrangians to reproduce the corresponding bouncing solution and the possibility of containing other important physical features to be considered a viable alternative to teleparallel gravity.

\section*{Acknowledgments}
AdlCD acknowledges financial support from projects 
FPA2014-53375-C2-1-P Spanish Ministry of Economy and Science, 
FIS2016-78859-P European Regional Development Fund and Spanish Research Agency (AEI), CA16104 COST Action EU Framework Programme Horizon 2020,
University of Cape Town Launching Grants Programme and National Research Foundation grants 99077 2016-2018, Ref. No. CSUR150628121624, 110966 Ref. No. BS170509230233 and the NRF Incentive Funding for Rated Researchers (IPRR), Ref. No. IFR170131220846.
DSCG is funded by the Juan de la Cierva programme (Spain) No. IJCI-2014-21733 and by MINECO (Spain), project FIS2016-76363-P. This article is based upon work from CANTATA COST (European Cooperation in Science and Technology) action CA15117, EU Framework Programme Horizon 2020. 


\end{document}